\documentclass[lettersize,journal]{IEEEtran}

\def\BibTeX{{\rm B\kern-.05em{\sc i\kern-.025em b}\kern-.08em
    T\kern-.1667em\lower.7ex\hbox{E}\kern-.125emX}}
\usepackage{balance}

\usepackage{amsmath,amssymb,amsfonts}

\usepackage[ruled,vlined,linesnumbered]{algorithm2e}
\usepackage{algpseudocode}
\usepackage{graphicx}
\usepackage{textcomp}
\usepackage{xcolor}
\usepackage{tcolorbox}
\usepackage{framed}
\usepackage{array}
\usepackage{multirow}
\usepackage{mathrsfs}
\usepackage{threeparttable}
\usepackage{subcaption}
\usepackage{enumitem}
\usepackage{float}
\usepackage{cases}
\usepackage{dutchcal}
\usepackage[]{hyperref} 
\usepackage{xspace}
\usepackage{cleveref}
\usepackage{hyperref}
\usepackage{listings}
\usepackage[colorinlistoftodos]{todonotes}
\usepackage{soul}
\usepackage{colortbl}
\usepackage{booktabs}

\usepackage{tikz}
\usetikzlibrary{shapes.geometric}

\tikzset{ellip/.style={ellipse, draw, thick,
    minimum height=3cm, minimum width=6cm, rotate=#1
}}

\newcommand{\ourapproachshort}{\texttt{LLMOrch}\xspace}
\newcommand{\ourapproach}{\ourapproachshort}

\newcommand{\ourgraph}{Function-call Relation Graph\xspace}
\newcommand{\ourgraphshort}{FRG\xspace}
\newcommand{\oururl}{\url{https://www.hostize.com/v/c3oLTBMUwn}\xspace}

\newcommand{\llmcompiler}{\texttt{LLMCompiler}\xspace}
\newcommand{\llmtoolcompiler}{\texttt{LLM-Tool Compiler}\xspace}

\definecolor{ColorCodeBackground}{RGB}{248,248,248}
\definecolor{ColorCodeKeyword}{RGB}{232,132,64}
\definecolor{ColorCodeComment}{rgb}{0,0.6,0}
\definecolor{ColorCodeString}{rgb}{0.58,0,0.82}
\definecolor{ColorCodeNumber}{rgb}{0.5,0.5,0.5}
\definecolor{ColorAlgoKeyword}{RGB}{0,0,255}
\definecolor{ColorAlgoComment}{RGB}{191,0,64}
\definecolor{ColorAlgoNumber}{rgb}{0.5,0.5,0.5}
\definecolor{ColorTableBackground}{rgb}{0.9,0.9,0.9}
\definecolor{ColorTableGood}{rgb}{0,0.6,0}
\definecolor{ColorTableBad}{cmyk}{0,0.90,0.86,0}

\setlist[itemize]{leftmargin=1.5em}
\setlist[enumerate]{leftmargin=1.5em}

\newcommand{\defaultcodefontsize}{\footnotesize}
\newcommand{\codefontsize}{\defaultcodefontsize}
\lstdefinestyle{defaultlststyle}{
  backgroundcolor=\color{ColorCodeBackground},
  basicstyle=\ttfamily\codefontsize,
  keywordstyle=\color{ColorCodeKeyword}\bf\ttfamily,
  commentstyle=\color{ColorCodeComment}\it,
  numberstyle=\tiny\color{ColorCodeNumber},
  stringstyle=\color{ColorCodeString},
  lineskip=-1pt,
  columns=fullflexible,
  keepspaces=true,
  breaklines=true,
  breakatwhitespace=false,
  captionpos=b,
  numbers=left,
  numbersep=4pt,
  xleftmargin=0pt,
  frame=tb,
  framerule=0.5pt,
  showspaces=false,
  basewidth=0.5em,
  showstringspaces=false,
  escapechar=§,
  showtabs=false,
  tabsize=2
}
\newcommand{\lstcode}[1]{\lstinline[basicstyle=\ttfamily\small]|#1|}
\newcommand{\txtcode}[1]{{\small\texttt{#1}}}

\lstdefinestyle{orchlststyle}{
  language=java,
  morekeywords={inout, comp, exec, iip, imdp, seq, summ, chksat},
}

\lstnewenvironment{orchprog}{
  \lstset{style=orchlststyle}
}{
  \lstset{style=defaultlststyle}
}

\lstdefinestyle{fseqstyle}{
  language=python,
  basicstyle=\ttfamily\footnotesize,
  breakatwhitespace=true,
  breakindent=6em,
  morekeywords={s1, s2, s3, s4, s5, s6, s7, s8, s9, s10, s11, s12, s13, search, python, self, chatbot, aes, math, wiki, join},
}

\lstnewenvironment{funcseq}{
  \lstset{style=fseqstyle}
}{
  \lstset{style=defaultlststyle}
}

\lstdefinestyle{promptlststyle}{
  basicstyle=\ttfamily\footnotesize,
  keywordstyle=\color{ColorCodeKeyword}\bf\ttfamily,
  breakatwhitespace=true,
  breakindent=0pt,
  lineskip=-1pt,
  numbers=none,
  numbersep=0pt,
  xleftmargin=6pt,
  framesep=1em,
  framexleftmargin=1em,
  morekeywords={Change, Summary, Thoughts, Category, File, Code, Suggestion}
}

\lstnewenvironment{userprompt}{
  \lstset{style=promptlststyle}
}{
  \lstset{style=defaultlststyle}
}

\newcommand{\boxedtext}[1]{
\begin{center}
    \begin{tcolorbox}[
        colback=gray!10!white,
        colframe=black,
        width=\linewidth,
        arc=0.5mm,
        auto outer arc,
        boxsep=0pt,
        boxrule=.5pt
    ]
        #1
    \end{tcolorbox}
\end{center}
}

\newcommand{\defaultalgofontsize}{\scriptsize}

\algrenewcommand{\algorithmiccomment}[1]{{\tiny\color{ColorCodeComment}// #1}}

\newcommand{\callfn}[2]{\txtcode{#1}\left(#2\right)}
\newcommand{\callinitfn}[2]{\txtcode{#1}\left\{#2\right\}}

\newenvironment{myalgorithm}{
\begin{algorithm}[tb]
  \defaultalgofontsize

  \DontPrintSemicolon

  \SetNoFillComment
  \SetKwSty{algokeywordsty}
  \SetFuncSty{algofuncsty}

  \SetKw{continue}{continue}
  \SetKwBlock{Thread}{thread}{}
  \SetKwProg{Fn}{function}{}{}
}{
\end{algorithm}
}

\newcommand{\smalltitle}[1]{\smallskip\noindent\textbf{#1}. }

\newcommand{\conglidrawbacks}[1]{}

\begin{document}

\title{Efficient Function Orchestration for Large Language Models}

\author{Xiaoxia Liu, Peng Di, Cong Li, Jun Sun, and Jingyi Wang*

\thanks{Xiaoxia Liu, Cong Li and Jingyi Wang* are with Zhejiang University, Zhejiang 310007, China (e-mail: liuxiaoxia@zju.edu.cn;
chifei.lc@antgroup.com; wangjyee@zju.edu.cn).}%
\thanks{Peng Di and Cong Li are with Ant Group, Zhejiang 310020, China (e-mail: dipeng.dp@antgroup.com;
chifei.lc@antgroup.com).}%
\thanks{Jun Sun is with Singapore Management University, Singapore 188065, Singapore (e-mail: junsun@smu.edu.sg).}%
}






\maketitle

\begin{abstract}
  Function calling is a fundamental capability of today's large language models, but sequential function calling posed efficiency problems. Recent studies have proposed to request function calls with parallelism support in order to alleviate this issue. However, they either delegate the concurrent function calls to users for execution which are conversely executed sequentially, or overlook the relations among various function calls, rending limited efficiency. This paper introduces \ourapproach, an advanced framework for automated, parallel function calling in large language models. The key principle behind \ourapproach is to identify an available processor to execute a function call while preventing any single processor from becoming overburdened. To this end, \ourapproach models the data relations 
(i.e., 
definition-use (def-use) dependencies
among different function calls and coordinates their executions by their control relations (i.e., mutual-exclusion) as well as the working status of the underlying processors. When comparing with state-of-the-art techniques, \ourapproach demonstrated comparable efficiency improvements in orchestrating I/O-intensive functions, while significantly outperforming (2$\times$) them with compute-intensive functions. \ourapproach's performance even showed a linear correlation to the number of allocated processors. We believe that these results highlight the potential of \ourapproach as an efficient solution for parallel function orchestration in the context of large language models.

\end{abstract}


\section{Introduction}
\label{sec:intro}

\IEEEPARstart{R}{ecent} advancements in large language models (LLMs)~\cite{achiam2023gpt, du-etal-2022-glm, touvron2023llama2} have led to the development of AI agents such as AutoGPT~\cite{autogpt}, SWE-agent~\cite{sweagent}, and Agentless~\cite{agentless}.
Besides programming tasks, LLM-driven agents also led to improvements in intricate real-world challenges, including scientific computations~\cite{romera2024mathematical, trinh2024solving}, software engineering~\cite{10440574, 10704582}, protein engineering~\cite{wang2023self,rapp2024self}, and cellular research~\cite{hao2024large}.

\begin{figure}[tb]
    \centering
    \includegraphics[width=\linewidth]{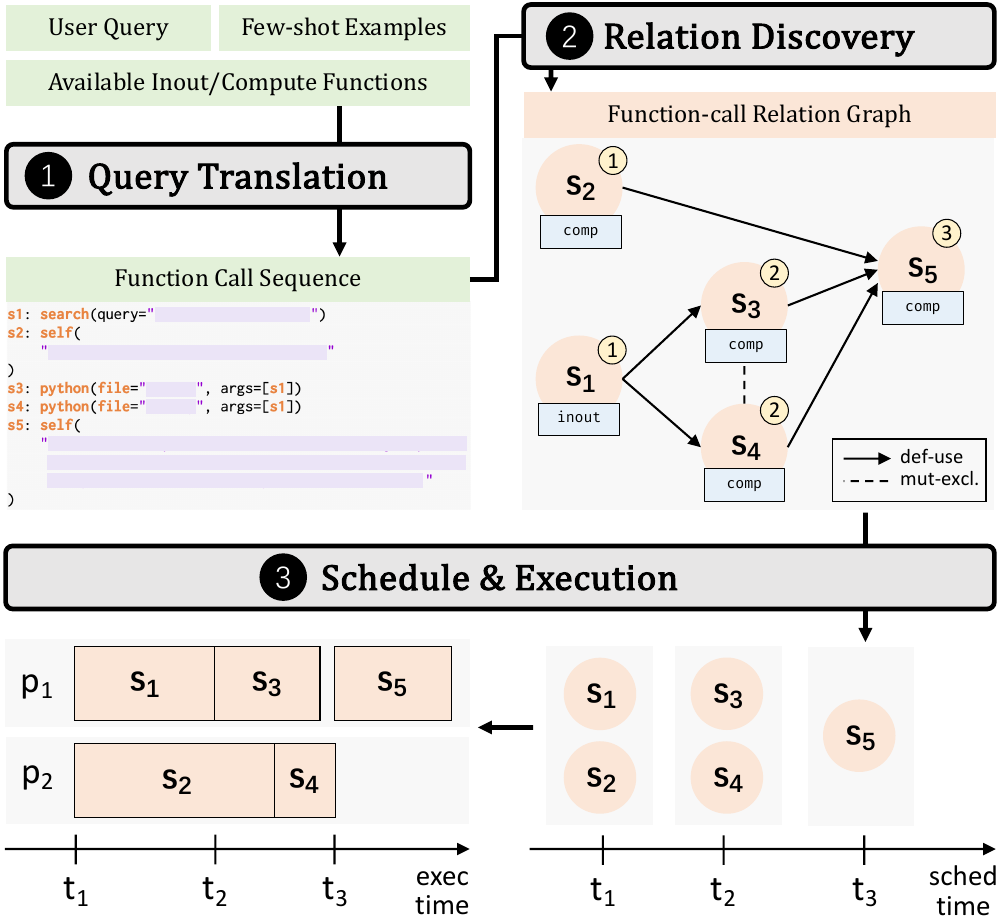}
    \caption{
        Overview of \ourapproach.
        Each node in the \ourgraph is assigned a rank, represented by a number in the top-right yellow circle, which is computed based on the \emph{def-use} (data) relations.
        The set of function calls with the same rank are scheduled concurrently for example $s_1$/$s_2$ and $s_3$/$s_4$, though their scheduling does not immediately trigger their execution.
        \ourapproach manages this coordination through their \emph{mutual-exclusion} (control) relations and the current work status of the underlying processors.
        In this example, $p_1$ and $p_2$ represent two physical processors;
        $s_3$ and $s_4$ are coordinated to them respectively because they are mutual-exclusive function calls.
    }
    \label{fig:overview}
\end{figure}

LLM-driven agents extensively rely on external user functions (or tools) to expand their capabilities.
To better align external functions with LLM's inherent capabilities, Yao et al. proposed the ReAct~\cite{yao2022react} framework for selecting functions and observing their outcomes in a sequential loop.
Since when, function calling has been one of the defacto, fundamental LLM capabilities.
For example, GPT-3.5~\cite{openaifunccall}, ChatGLM3~\cite{chatglm3}, and LLaMA3~\cite{LLaMa3} support native function calling.
LangChain~\cite{Chase_LangChain_2022} supplies a rich and convenient toolbox for a wide range of LLMs to call user functions.
There are also datasets~\cite{glaiveai,patil2023gorilla} and tools~\cite{qin2023toolllm} for augmenting existing LLMs with function calling capabilities.
All these works significantly benefit LLM's capability towards function calling.
Yet, their sequential nature poses a considerable problem towards efficiency~\cite{dao2023flashattention2, kwon2023efficient, fu2024break, zheng2023efficiently}, 
as well as software engineering (SE) concerns including maintainability, fault tolerance, and system predictability—key challenges in building reliable LLM-driven software systems.
To alleviate the problem, recent studies focus on calling functions concurrently.
OpenAI~\cite{openaipfcrel,openaipfcpr,openaipfcandex1} and Google~\cite{geminipfc} released Parallel Function Calling (or ParallelFC) for empowering their own GPT-/Gemini-series LLMs to request multiple function calls in one LLM response.
\llmtoolcompiler~\cite{singh2024llm} further accelerates the execution by fusing related operations into a single call, enabling token reuse and latency reduction.
\llmcompiler, on the other hand, is able to concurrently execute function calls for \emph{any} LLMs, even without ParallelFC capability~\cite{kim2024an}.
These works have achieved modest efficiency improvements when addressing user queries. 
However, they primarily focus on performance optimizations and lack a structured SE perspective—they do not provide a principled framework for managing dependencies, coordinating resources, or handling failures, which are essential for the design, maintenance, and evolution of production-grade LLM-based software.

Despite their efforts, we identified two major limitations.
(1) \emph{Lack} of automatic parallel orchestration: Neither ParallelFC nor \llmtoolcompiler supports automatic parallel orchestration, necessitating manual scheduling and execution by users.
This limitation has resulted in sequential processing of function calls, even in those ParallelFC's official examples~\cite{openaipfcandex1,openaipfcex2}.
(2) \emph{Lack} of efficient parallel orchestration:
While \llmcompiler attempts to improve parallelization performance by managing the schedule and execution of function calls, it often fails to consider the relations (either dependencies or exclusions) among function calls, rendering only limited efficiency improvements.
These limitations present challenges for systems with resource constraints that depend on LLM-driven agents, for instance, AI-powered operating systems~\cite{mei2024aios,pmlr-v235-wang24h} and robotic systems~\cite{lami,saycan,gao2024dagplan}.
In these systems, the core LLM-drivens agent must organize and execute numerous (I/O- or compute-intensive) tasks such as image or text processing, data analysis, and robotic arm movement with limited computational resources.
Improper function orchestration can lead to degraded performance.

In program analysis, a Definition-Use (def-use) chain~\cite{10.5555/1177220} describes the relationship between the point where a variable is defined (assigned a value) and the points where that value is used. We adapt this concept to model data flow between function calls.

\smalltitle{\ourapproach}
We present \ourapproach (\Cref{fig:overview}), an automatic and efficient function orchestrator with fine-grained parallelism support for parallel function calling for large language models.
The key idea behind \ourapproach is to identify an available processor to execute a function call while preventing any single processor from becoming overburdened.
In this paper, ``over-burden'' or ``overwhelm'' refers to a processor being assigned more work than it can handle efficiently, leading to queuing delays, increased latency, or potential stalls.
To achieve this, we design \ourapproach around two key features:
(1) Discover data and control relations from function calls and (2) Separate function call scheduling and execution to enable better parallelism.

In this paper, data relations (i.e., \emph{def-use}) refer to the data flow dependencies among various function calls, while control relations (i.e., \emph{mutual-exclusion}) model their execution (I/O- or compute-intensive) conditions.
\ourapproach's function orchestration follows a work-sharing style~\cite{workstealing} in multithreaded computation, yet it differentiates between these two types of relations.
In particular, for a user query and all available functions, \ourapproach builds a \emph{\ourgraph} (\ourgraphshort) to represent the relations, discovered from a function call sequence--a sequential, step-by-step plan for addressing the user query through function calls.
\ourapproach then follows \ourgraphshort to schedule function calls concurrently and coordinate their executions.
Specifically, a function call is scheduled for execution immediately after all its data-dependent function calls are scheduled, this 
allows
for multiple function calls to be scheduled at a time.
However in \ourapproach, the scheduling does not lead to their direct execution.
Instead, \ourapproach avoids the concurrent execution of multiple mutual-exclusive function calls on one processor.
\ourapproach queues the set of concurrent function calls and distributes them to execute on specific processors, if the processors are available and their mutual-exclusive calls are already distributed (either to a different processor or the same one after they complete).



When compared against ReAct~\cite{yao2022react}, ParallelFC~\cite{openaipfcandex1}, and \llmcompiler~\cite{kim2024an}, \ourapproach demonstrated comparable efficiency improvements in orchestrating I/O-intensive functions, while significantly outperforming them with functions involving considerable computations.~\conglidrawbacks{The evaluation can be stronger.}
Moreover, we observed that the improvements achieved by \ourapproach positively co-related to the number of allocated processors.
In real-world scenarios ``Purchase Intent Analysis'' and ``End-to-End Encryption'', \ourapproach successfully addressed the given user queries with satisfactory answers similar to prior works.
We believe that these results highlight \ourapproach's usefulness and practicability, which we contribute to the discovery of data and control relations, as well as the separation of function call scheduling and execution.
Furthermore, we hope that \ourapproach represents a small yet significant advancement in the field of function orchestration.

\smalltitle{Contributions} 
Our main contributions are:
\begin{itemize}
    \item We present a framework, \ourapproach, designed to orchestrate function calls in parallel by introducing the novel consideration of both data and control relations, and separating the processes of scheduling and execution.
    \item We evaluate \ourapproach against state-of-the-art techniques where \ourapproach achieved superior performance when orchestrating both I/O- and compute-intensive functions.
    \item We open-source \ourapproach to facilitate the community and future research: \oururl.
\end{itemize}

\section{Illustrative Example}\label{sec:illustrative_example}

\ourapproach decomposes the orchestration of function calls for an LLM into three steps as illustrated in \Cref{fig:overview}:
(1) Query Translation,
(2) Relation Discovery, and
(3) Schedule \& Execution.
This section walks through the process following an example with a (simplified) user query:
\begin{userprompt}
I want to do a search engine optimization (SEO) audit. I'll firstly search the webpage (https://openai.com/). I also need chatbot for a SEO optimization example. Then I'll request the local python file ("seo.py") for Technical SEO Audit, and the local python file ("con.py") for Content Analysis. Finally, I prefer a summary report.
\end{userprompt}
and available functions:
\begin{itemize}
    \item \lstcode{search(query)}: Search \lstcode{query} through search engines and obtain the top-k results.
    \item \lstcode{chatbot(prompt)}: Ask an LLM for an answer or a completion towards \lstcode{prompt}.
    \item \lstcode{python(file,args)}:
    Execute the python code in \lstcode{file} with arguments \lstcode{args}.
\end{itemize}
In this example, we will orchestrate these functions for GPT-4.
We deployed LLaMA3-8B locally on CPUs (i.e., neither on GPUs nor remotely via a remote API call)%
\footnote{
This can be accomplished via like llama.cpp~\cite{llamacpp}, MLC-LLM~\cite{mlcllm}, PowerInfer~\cite{powerinfer}, etc.
The scenario is realistic and common for advanced users today and we believe it is the future for mobile operating systems~\cite{powerinfer2,llminaflash}.
}
for \txtcode{chatbot}.

\begin{figure}[tb]
    \centering
    \input{figures/fseq_example}
    \caption{
        The function call sequence of our illustrative example after query translation.
        The grammar is similar to those used by ReAct and \llmcompiler.
        Each function call is given a unique ID (e.g., \lstcode{s1}) which also serves as the result of the function call.
        Different function calls have explicit data-dependencies and implicit control exclusions;
        these decide their order for subsequent schedule and execution.
    }
    \label{fig:example-fseq}
\end{figure}

\smalltitle{Query Translation}
Given the user query and all available functions, \ourapproach instructs LLMs to translate the query into a sequence of function calls.
Unlike the defacto function call~\cite{lund2023chatgpt} and the ReAct~\cite{yao2022react} paradigm where LLMs request function calls progressively, we require LLMs to comprehend the user query and organize a function call sequence, following \llmcompiler~\cite{kim2024an} and AutoGPT~\cite{autogpt}.
We demand LLMs to attach each function call with a \emph{unique ID}.
For function calls requiring results of prior calls, the unique IDs (or IDs) serve as their results and can be referred to as arguments.

\Cref{fig:example-fseq} displays the function call sequence for our illustrative example, by GPT-4~\cite{achiam2023gpt}.
In this example, GPT-4 first conducted search using \lstcode{search}, then prompted LLaMA3 for an SEO example.
It then sequentially executed \lstcode{seo.py} and \lstcode{con.py} using the results obtained in \lstcode{s1} respectively for technical SEO and content analysis, and finally summarized a report by LLaMA3.

\smalltitle{Relation Discovery}
\ourapproach discovers the inter-relations among various function calls.
In this paper, we consider two types of relations: data relations and control relations.

Data relations refer to the data flow relations among different function calls.
Such relations are indicated by the definition of a call's results and all its uses in other calls' arguments.
In the function call sequence, we explicitly capture these \emph{def-use relations} through unique IDs and their references.
The function calls, together with their \emph{def-use relations}, construct a data dependency graph which is directed and acyclic.
Accordingly, we assign each function call with a \emph{rank} following their topological order in the graph.
For instance, function calls \lstcode{s3} (rank: 2) and \lstcode{s4} (rank: 2) are data-dependent (in solid arrows) on the function call \lstcode{s1} (rank: 1) in \Cref{fig:overview}.
It is worthwhile mentioning that function calls with the same rank means they can be called concurrently.~\conglidrawbacks{Not exactly, they can be scheduled even if they are not. So this is a sufficient condition, not necessary.}

Control relations model the behaviors of the functions.
In this paper, we categorize available functions into two traditional groups:
I/O-intensive functions (or \emph{inout} functions) that primarily execute input-output operations without tying up processors, and
compute-intensive functions (or \emph{compute} functions) that exclusively occupy a processor until they pause or stop.
As the first work that attempts to analyze control relations in parallel function orchestration, we conservatively consider that two function calls are \emph{mutual exclusive} to each other if both of them are compute functions with the same rank.
In our example, \lstcode{search} is an inout function and \lstcode{chatbot} and \lstcode{python} are compute functions.
is mutual
exclusive (in dashed lines) to \lstcode{s4} (rank: 2).

We incorporate all identified control relations into the data dependency graph by introducing ``bidirectional, virtual edges''.
In this paper, we refer to the resulting directed acyclic graph as \emph{\ourgraph} (\ourgraphshort).
The function call sequence is scheduled and subsequently executed according to \ourgraphshort.
The \ourgraphshort for \Cref{fig:example-fseq} is presented in \Cref{fig:overview}.

\smalltitle{Schedule \& Execution}
\ourapproach schedules (or submits) function calls for execution based on their rank and data relations in \ourgraphshort, starting with calls assigned a rank of 1.
Note that scheduling a function call does not immediately result in its execution; \ourapproach coordinates its execution with other function calls to achieve commendable parallelism.
As a completed function call becomes available, \ourapproach concurrently submits all function calls data-dependent on it and with an incremental (+1) rank.
The scheduling process stops once all function calls have been scheduled.

As for the \ourgraphshort of our illustrative example (\Cref{fig:overview}), the function calls \lstcode{s1} and \lstcode{s2} are initially scheduled concurrently.
Upon completion of \lstcode{s2}, \lstcode{s5} will not be scheduled until \lstcode{s3} and \lstcode{s4}--which are scheduled after \lstcode{s1}'s completion--also finish.

For each set of concurrent function calls scheduled for execution, \ourapproach queues them and then coordinates them according to their control relations in \ourgraphshort and the work status of the underlying processors.
The principle is to find a spare processor to execute a function call while preventing any processor from becoming overburdened (\Cref{algo:exec-coord}).
The coordinator thereby prioritizes the execution of inout function calls and separates the execution of mutual-exclusive calls;
If there is an insufficient number of processors, \ourapproach sequentializes them.

Following the \ourgraphshort in \Cref{fig:overview}, \lstcode{s1} is executed before \lstcode{s2} if there is only one processor available.
\lstcode{s3} and \lstcode{s4} will be executed on separate processors if there are over two spare processors, otherwise they may be executed sequentially.

\begin{table}[tb]
  \newcommand{\yes}{\color{ColorTableGood} $\checkmark$}
  \newcommand{\nop}{\color{ColorTableBad} $\times$}
  \newcommand{\ltcompiler}{\texttt{LTCompiler}\xspace}

  \centering
  \footnotesize
  \setlength{\tabcolsep}{.1em}

  \caption{
    Comparison with state-of-the-art works on function calling and orchestration.
    No works except \ourapproach consider control relations and support execution coordination. 
    \ltcompiler represents \llmtoolcompiler, featured with the ability of function fusing. 
  }
  \label{tab:comp-sota}

  \begin{tabular}{lcccccc}
    \toprule
    \multirow{2}{*}{\textbf{Work}}       & {Sequential} & {Parallel} & {Data}     & {Control}  & {Execution} & {Errors}\\
                                         & {Calling}   & {Calling} & {Relation} & {Relation} & {Coordina.} & {Recovery}\\
    \midrule
    ReAct~\cite{yao2022react}            & \yes         &  \nop      & \nop       & \nop       & \nop    & \nop    \\
    ParallelFC~\cite{openaipfcandex1}    & \yes         &  \yes      & \nop       & \nop       & \nop    & \nop    \\
    \ltcompiler~\cite{singh2024llm}      & \yes         &  \yes      & \nop       & \nop       & \nop     & \nop   \\
    \llmcompiler~\cite{kim2024an}        & \yes         &  \yes      & \yes       & \nop       & \nop    & \nop    \\
    \ourapproach                         & \yes         &  \yes      & \yes       & \yes       & \yes     & \yes   \\
    \bottomrule
  \end{tabular}
\end{table}

\smalltitle{Discussions}
We recognize that existing works ReAct~\cite{yao2022react}, ParallelFC~\cite{openaipfcandex1}, and \llmcompiler~\cite{kim2024an} can accomplish the task with satisfactory results.
However, they all struggle when it comes to fast function orchestration in resource-constrained settings.
For instance, \ourapproach is able to provide responses to the user query in our illustrative example in 53.22 seconds when running on a device with two physical processors.
However, the state-of-the-art \llmcompiler took 81.32 seconds to accomplish the same query, while ParallelFC and ReAct required 85.28 and 93.37 seconds, respectively.
ReAct's unsatisfactory efficiency is attributed to its sequential nature.
ParallelFC is always waiting for the completion of all parallel function calls before proceeding with the next call.
\llmcompiler directly executes all concurrent function calls (like \lstcode{s2}, \lstcode{s3}, and \lstcode{s4}) after each completed function call (like \lstcode{s1}), without considering the control relations among them and without providing flexible execution coordination.
In such cases, a system may become stuck if there are too many compute function call processes.
In contrast, \ourapproach delays the execution of a compute function call unless there are processors available.
This significantly sets our work apart from all the others.
We summarize the key differences of these works in \Cref{tab:comp-sota}.

\section{The \ourapproach Framework}

This section expands \Cref{sec:illustrative_example} with implementation details.

\subsection{Query Translation}

\ourapproach begins by translating the user query into a \emph{Function Call Sequence}, following the context-free grammar provided below, which also serves as annotations for this section:
\conglidrawbacks{The expression calculations are overclaims.}

\begin{figure}[h]
    \centering

    \footnotesize
    \setlength{\tabcolsep}{.8em}

    \newcommand{\tinyarrow}{{\tiny \txtcode{->}}}
    \newcommand{\widehspace}{\hspace{.3em}}
    \newcommand{\widemid}{\widehspace\mid\widehspace}

    \framebox{\begin{tabular}{lccl}
        Sequence & $\sigma$ & $\to$ & 
            $\kappa^+$ \\
    
        Func Call & $\kappa$ & $\to$ & 
            $i\txtcode{:}~f\txtcode{(}\alpha^*\txtcode{)}$ \\
    
        Unique ID & $i$ & $\to$ & 
            $\txtcode{s}_0$ $\widemid$ 
            $\txtcode{s}_1$ $\widemid$ 
            $\txtcode{s}_2$ $\widemid$ 
            $\cdots$ \\
    
        Argument & $\alpha$ & $\to$ & 
            $n = e$ \\

        Expression & $e$ & $\to$ &
            $i$ $\widemid$
            $v$ $\widemid$
            $e\txtcode{+}e$ $\widemid$
            $e\txtcode{-}e$ $\widemid$
            $\cdots$ \\

        Arg Name & $n$ & $\to$ & 
            name of a function argument \\
        Arg Value & $v$ & $\to$ &
            number $\widemid$ 
            string $\widemid$ 
            array $\widemid$
            $\cdots$ \\
    
        Function & $f$ & $\to$ & 
            $\txtcode{self}$ $\widemid$
            user-defined functions
    \end{tabular}}
\end{figure}

This grammar was designed to be in line with ReAct~\cite{yao2022react} and \llmcompiler~\cite{kim2024an}, with the intention of making the generated programs simple, easily learned by LLMs.
In particular, a function call sequence ($\sigma$) is an array of function calls ($\kappa$), each with a unique ID ($i$) assigned to it.
These unique IDs can be referenced to by other function calls as arguments ($\alpha$).
This establishes the def-use relations---the data relations that we consider in this paper---for scheduling them.
The functions ($f$) that can be included in a function call sequence are either \lstcode{self} or those defined by users to address the user query.
It is important to note that the sequences generated are indifferent to inout and compute functions.
However, \ourapproach is sensitive to them, and we ask users to explicitly label them as either inout or compute when passing them to \ourapproach.
\conglidrawbacks{It looks a little bit weak: we manually label them; how about intelligently recognizing them?}
To clarify, ``self'' is a reserved built-in function that instructs the LLM to leverage its inherent capabilities (e.g., reasoning, knowledge recall, text generation) for task execution, instead of invoking external tools or user-defined functions.

To facilitate translation, we crafted prompts with few-shot examples, which we found to be more effective than directly prompting LLMs with the grammar.
We conducted a small study involving 100 user queries from the HotpotQA~\cite{yang2018hotpotqa} benchmark, of which over 80\% failed when generating with the grammar.
Conversely, the successful ratio with few-shot examples is over 85\%.
\conglidrawbacks{I suppose 80\% is an upper bound for using the grammar and 85\% is a lower bound for few-shot examples.}


\subsection{Relation Discovery}

\ourapproach then builds a \emph{\ourgraph} (\ourgraphshort) to represent the data and control relations among function calls. 
\smalltitle{\ourgraph}
The \ourgraphshort of a function call sequence $\sigma = [\kappa_1, \kappa_2, \cdots, \kappa_{|\sigma|}]$ is a directed acyclic graph $G = \left\langle K, R_1, R_2 \right\rangle$:
\begin{itemize}
    \item $K = \left\{\kappa \mid \kappa \in \sigma \right\}$ is the set of function calls in the sequence $\sigma$, representing $G$'s nodes.
    \item $R_1: K \times K$ is the set of def-use (data) relations among $K$, serving as $G$'s directed edges:
    $$
        \hspace{2em}
        \forall \langle\kappa', \kappa''\rangle \in R_1.~
        \exists\alpha.~
        \callfn{uses}{\kappa''.\alpha.e,~\kappa'.i}
    $$
    where $\callfn{uses}{e,i}$ determines whether the expression $e$ uses the definition $i$.
    \item $R_2: K \times K$ is the set of mutual-exclusion (control) relations.
\end{itemize}

\noindent
$R_2$ represents mutual-exclusion (control) relations among function calls. We define it in two levels:

\vspace{1.5mm}
\noindent
\textbf{\textit{General Resource-Contended Definition.}} Each function call $\kappa$ has a resource profile $\rho(\kappa) \in \mathbb{R}_{\ge 0}^{|\mathcal{T}|}$, where $\mathcal{T} = \{\text{CPU}, \text{GPU\_MEM}, \text{IO}, \text{MEM}, \dots \}$ denotes resource types. The system has a global capacity vector $\mathbf{C}$. The Mode Flag $\lambda(\kappa) \in \{\mathsf{BLOCK}, \mathsf{NONBLOCK}\}$ for I/O tasks, specifying whether the I/O operation is blocking or non-blocking. Two function calls $\kappa'$ and $\kappa''$ are mutually exclusive if there exists $i \in \mathcal{T}$ such that
\[
    \rho_i(\kappa') + \rho_i(\kappa'') > \mathbf{C}_i.
\]
This formulation naturally supports hybrid tasks that simultaneously require multiple resources (e.g., CPU + GPU + I/O).

\vspace{1.5mm}
\noindent
\textbf{\textit{Simplified Model for Evaluation.}} Although the general resource contention model can capture hybrid tasks that consume multiple resources, we adopt a simplified binary model for practical evaluation, distinguishing compute and I/O tasks. Specifically, each function call $\kappa$ has a resource profile  $\rho(\kappa) \in \mathbb{R}_{\ge 0}^{|\mathcal{T}|}$ with $\mathcal{T} = \{\text{CPU}, \text{MEM}, \text{GPU\_MEM}, \text{IO}\}$: 
\begin{itemize}
    \item For a compute function $\kappa_{\text{comp}}$, 
    $\rho(\kappa_{\text{comp}}) = (1, \delta, 0, \text{LOW})$
    \item For a blocking I/O function $\kappa_{\text{IO}}$,
    $\rho(\kappa_{\text{IO}}) = (\epsilon, \delta', 0, \text{HIGH}), \quad \lambda(\kappa_{\text{IO}}) = \text{BLOCK}$
    \item For a non-blocking I/O function $\kappa_{\text{IO}}$,
    $\rho(\kappa_{\text{IO}}) = (\epsilon, \delta'', 0, \text{HIGH}), \quad \lambda(\kappa_{\text{IO}}) = \text{NONBLOCK}$
\end{itemize}
Here, $\epsilon$ is negligible CPU usage. This simplified model serves as a concrete instantiation of the general framework.
Therefore, for any $\left(\kappa', \kappa''\right) \in R_2$:
    $$
        \hspace{2em}
        \callfn{comp}{\kappa'.f} \land
        \callfn{comp}{\kappa''.f} \land
        \callfn{rank}{\kappa'} = \callfn{rank}{\kappa''}
    $$
where $\kappa.f$ refers to the function invoked by the function call $\kappa$, and 
$\callfn{comp}{f}$ decides if the function $f$ is a compute function;
$\callfn{rank}{\kappa}$ returns the \emph{rank} of a function call $\kappa$.

\begin{myalgorithm}
  \renewcommand{\txtcode}[1]{\defaultalgofontsize \texttt{#1}}

  \caption{Assigning ranks}\label{algo:assign-rank}
  
  \Fn{Assigner$(\mathrm{FnCalls}~K, \mathrm{DataRelations}~R_1)$}{
    $R \gets R_1$\\
    $Q \gets \callinitfn{orderedset}{\kappa'' \in K \mid \nexists \kappa' \in K.~\left\langle \kappa', \kappa'' \right\rangle \in R }$\\
    \lFor{\rm $\textnormal{each}~\kappa \in Q$}{
        $\callfn{setrank}{\kappa, 1}$
    }
    \While{\rm $|Q| \neq 0$}{
        $\kappa' \gets \callfn{popfront}{Q}$\\
        \For{\rm $\textnormal{each}~\kappa'' \in K.~\langle\kappa', \kappa''\rangle \in R$}{
            $R \gets R~~/~~\left\{\langle \kappa', \kappa'' \rangle\right\}$\\
            \If{\rm $\nexists \kappa''' \in K.~\left\langle \kappa''', \kappa'' \right\rangle \in R$}{
                $\callfn{setrank}{\kappa'', \callfn{rank}{\kappa'} +1}$\\
                $\callfn{pushback}{Q, \kappa''}$
            }
        }
    }
  }
\end{myalgorithm}



\smalltitle{Rank Assignment}
While constructing \ourgraphshort, specifically after creating $K$ and $R_1$ and before creating $R_2$, \ourapproach assigns each function call $\kappa \in K$ with a \emph{rank} based on  their topological order as defined by $K$ and $R_1$.
In particular, the \emph{Rank Assigner} (\Cref{algo:assign-rank}) starts from nodes without incoming edges (i.e., function calls without data dependencies, Lines~3--4).
We iteratively assign a function call $\kappa''$ with an incremental rank (Line~10) if it is data-dependent on $\kappa'$ (Line~7) and all other data dependencies have been considered (Line~9).
The process stops when all function calls have been assigned a rank (Line~5).

\smalltitle{Discussion}
Prior works~\cite{kim2024an} consider the def-use relations among different function calls while orchestrating them like ours.
However, they did overlook their control relations, which were shown to be effective for efficient function orchestrations (\Cref{ssec:evaluation_compute}).
In this paper, we consider the mutual-exclusive control relations and combine them with the def-use data relations into \ourgraphshort.
It guides \ourapproach's subsequent function orchestration.

\begin{myalgorithm}
  \renewcommand{\txtcode}[1]{\defaultalgofontsize \texttt{#1}}

  \caption{Scheduling function calls}\label{algo:func-sched}
  
  \Fn{Scheduler$(\mathrm{FnCalls}~K, \mathrm{DataRelations}~R_1)$}{
    $S \gets \callinitfn{set}{}$ \;
    $\callfn{sendcoord}{\left\{\kappa \in K \mid \callfn{rank}{\kappa} = 1 \right\}}$\;
    \While{\rm $|S| \neq |K|$}{
        $C \gets \callfn{recvcoord}{}$\;
        \While{\rm $|C| \neq 0$}{
            $\kappa' \gets \callfn{popfront}{C}$\;
            $S' \gets \left\{ \kappa'' \in K \mid \right.$
            $\left. \quad \langle \kappa', \kappa'' \rangle \in R_1 \land \callfn{rank}{\kappa''} = \callfn{rank}{\kappa'}+1 \right\}$\;
            $\callfn{sendcoord}{S'}$\;
            $S \gets S \cup S'$\;
        }
    }
  }
\end{myalgorithm}




\subsection{Schedule \& Execution}

\ourapproach follows a work-sharing style~\cite{workstealing} to schedule and execute the function call sequence $\sigma$ based on its \ourgraphshort: $G = \langle K, R_1, R_2 \rangle$.
Unlike prior works~\cite{kim2024an,openaipfcandex1}, \ourapproach separates the schedule of a function call from its execution to achieve a better parallelism.
The basic idea is that:
A function call will be scheduled once all its preceding data-dependent (as indicated by $R_1$) function calls are complete, while it may not be immediately executed;
\ourapproach coordinates the execution of all submitted, concurrent function calls according to their mutual-exclusion relations (as implied by $R_2$) and the work status of the underlying processors.

\ourapproach thereby involves a \emph{Call Scheduler} and an \emph{Execution Coordinator}.
They interact with each other through the following interfaces:
\begin{itemize}
    \item $\callfn{sendcoord}{S}$:
    The scheduler submits to the coordinator a set $S$ comprising function calls intended to be executed concurrently.
    
    \item $\callfn{sendsched}{C}$:
    The coordinator sends back to the scheduler the set of all completed calls $C$ whenever new completions are available.

    \item $\callfn{recvcoord}{}$:
    The scheduler waits until the coordinator returns all completed function calls.
    
    \item $\callfn{recvsched}{}$:
    The coordinator waits until the scheduler provides a new set of function calls for execution.
\end{itemize}

\smalltitle{Call Scheduling}
The scheduler (\Cref{algo:func-sched}) submits function calls for coordination and execution based on their data relations (i.e., def-use relations as reflected by their ranks in this paper)--a common practice in task scheduling~\conglidrawbacks{Cite some task scheduling paper; they typically determine the order by a DAG}.
The scheduler begins by submitting all function calls with rank 1 (Line~4) as they have no dependencies.
The scheduler then waits until one of them completes, notified by the coordinator (Line~7).
Upon the completion of a function call $\kappa'$, the scheduler identifies the next set $S'$ of concurrent calls and submits them for execution (Lines~9--12).
Each function call $\kappa'' \in S'$ should be data-dependent on $\kappa'$ and have an incremental rank (Line~10), meaning that it is data-dependent solely on $\kappa'$ after $\kappa'$ completes.
The process continues until all function calls have been scheduled (Line~5).

\begin{myalgorithm}
  \renewcommand{\txtcode}[1]{\defaultalgofontsize \texttt{#1}}

  \caption{Coordinating function call executions}\label{algo:exec-coord}
  
  \Fn{Coordinator$(\mathrm{FunctionCalls}~K, \mathrm{ControlRelations}~R_2)$}{
    $S \gets \callinitfn{deque}{}$ \Comment{Function calls that have been scheduled}\; 
    $C \gets \callinitfn{set}{}$ \Comment{Function calls that have completed}\; 

    \Thread{
        \While{$|S| \neq |K|$}{
           ~\Comment{Wait for scheduler sending back newly scheduled function calls}\;
            $\callfn{pushback}{S, \callfn{recvsched}{}}$\;
        }    
    }    

    \While{\rm $|C| \neq |K|$}{
        $P \gets \callfn{spareprocs}{}$
        \Comment{Wait until there are spare processors}\;
        $C' \gets \textnormal{the~ordered~set~of~calls~ever~completed~on~} P$\;
        $C \gets C \cup C'$\;
        ~\Comment{Notify scheduler of newly completed function calls}\;
        $\callfn{sendsched}{C'}$\;
        $S' \gets \callfn{popfront}{S}$
        \Comment{Fetch the first set of concurrent calls for coordination and execution}
        \;
        $M \gets \left\{ \kappa' \in S' \mid \exists \kappa'' \in S'.~\left( \kappa', \kappa'' \right) \in R_2 \right\}$\;
        $I \gets S'~~/~~M$\;
        \If(\Comment{Inout functions are given higher priorities}){$|I| \neq 0$}{
            $\callfn{execute}{I, \left\{ \callfn{random}{P} \right\}}$ \Comment{Distribute all inout function calls to processor $p$}\;
        }
        \If(\Comment{Execute them either exclusively or sequentially}){$|M| \neq 0$}{
            $\callfn{execute}{\callfn{popfront}{M, \txtcode{count=}|P|}, P}$\;
            \lIf{$|M| \neq 0$}{
                $\callfn{pushfront}{S, M}$
                \Comment{Re-coordinate the remaining function calls in the next iteration}
            }
        }
    }
  }
\end{myalgorithm}

\smalltitle{Execution Coordination}
Not all submitted function calls are executed immediately.
The coordinator's principle is to map function calls to available processors based on their resource profiles $\rho(\kappa)$ and the system's capacity $\mathbf{C}$, preventing any resource from becoming overburdened.
To this end, the coordinator enforces constraints based on control relations (i.e., the resource profiles $\rho(\kappa)$) and the current work status of the underlying processors:
\begin{enumerate}
    \item Mutual-exclusive calls should not be coordinated to the same processor for execution as each of them occupies the processor until completion.
    \item Inout functions are given higher priorities than computing functions as they will be tied up. Blocking I/O calls, i.e., $\lambda(\kappa)=\mathsf{BLOCK}$, are prioritized to exploit their negligible CPU usage ($\rho_{\mathsf{CPU}} \approx 0$), quickly freeing the processor to overlap computation with I/O latency. Non-blocking calls, i.e., $\lambda(\kappa)=\mathsf{NONBLOCK}$, are offloaded to a dedicated \textit{Async I/O Pool}, preventing the coordinator bottleneck by handling them under separate resource constraints. 
    \item If there is an insufficient number of spare processors, \ourapproach can break the first constraint.
\end{enumerate}
Here, this \textit{Async I/O Pool} manages execution against the $\mathbf{C}_{\mathsf{IO}}$ constraint using an event loop, preventing the central coordinator from becoming a bottleneck for high-throughput I/O operations.

\conglidrawbacks{ Looks like there're some inconsistencies with your implementations like I have coordinates all i/o functions to a single processor. Please check them and tell me if my writings are appropriate. If  not, please let me know asap.}
The detailed process is presented in \Cref{algo:exec-coord}.
The coordinator starts by initializing a thread to receive and queue newly scheduled function calls (Lines~4--7).
When a set of available processors $P$ (Line~9) is present, the coordinator notifies the scheduler of all the completed function calls (Lines~10--13).
Subsequently, the coordinator coordinates the execution of the earliest set $S'$ of scheduled, concurrent function calls (Line 14).
In particular, it identifies all mutual-exclusive compute function calls ($M$, Line 15) and all the remaining inout function calls ($I$, Line~16).
The coordinator prioritizes by assigning all inout calls to a single processor as each will be tied up after execution (Lines~17--18).
As for mutual-exclusive calls, the coordinator separates their executions by fetching the first $|P|$ calls and distributing each to a single available processor (Lines~19--20).
If there are any remaining calls yet to be coordinated, the coordinator manages them in the next iteration (Line~21) such that the processors are not overburdened.

\smalltitle{Discussion}
In addition to control relations, the separation of function call scheduling and execution as well as the coordination of function call executions distinguish our work from all previous ones~\cite{yao2022react,openaipfcandex1,kim2024an}.
Even though \ourapproach has employed a simple coordination policy (following work-sharing~\cite{workstealing}) when realizing our instinctive principle, the coordinator was evaluated to be effective when providing an efficient function orchestration, especially for compute calls, as shown in \Cref{ssec:evaluation_compute}.
In contrast, \llmcompiler overlooks the availability of computational resources, directly delegating function call execution to the underlying runtime's scheduler, specifically Python's co-routine scheduler \txtcode{asyncio}~\cite{pyasyncio}, with all function calls being executed in a single thread.
While the principle of distributing mutually exclusive tasks is conceptually straightforward, its implementation necessitates a stateful coordinator that resolves the fundamental tension between static planning, dynamic resource availability, and execution constraints. In practice, we formalize these control relations through a Function-call Relation Graph (FRG), which requires accurate classification of heterogeneous LLM-generated functions. Moreover, implementing dynamic scheduling on a ``C+MPI'' architecture adds further complexity in asynchronous message passing, processor coordination, and fault recovery. These factors underline that the proposed technique is far from trivial and directly underpin the significant efficiency gains reported in our evaluation.



\conglidrawbacks{Re-implement the overall work and expose below interfaces}


\subsection{Errors \& Recovery}

\ourapproach is likely to encounter various errors.
To address this, it incorporates a recovery mechanism to handle two types of errors: 
(1) Compile-time errors, which arise when LLMs fail to translate the user query into a valid function call sequence, resulting in such as invalid syntax and undefined IDs;
(2) Runtime errors, which occur when the execution of a function call does not normally exit.

\smalltitle{Compile-time Recovery}
\ourapproach employs a feedback-driven method---a standard treatment in the current state of program repair~\cite{chatRepair} and LLM-driven agents~\cite{metamut,autogpt}---to manage compile-time errors.
When it fails to parse a function call sequence due to syntax or semantic violations, \ourapproach creates a repair prompt containing the error message, the erroneous function call sequence, and the few-shot examples used for user query translation.
It then requests an LLM to repair the function call sequence to align with the provided examples.
This repair process continues until no compile-time errors are detected or a maximum number of attempts is reached.
If the repair fails, \ourapproach increases the temperature and calls the LLM to re-translate the user query from scratch, allowing for a limited number of additional attempts.
If all attempts fail, \ourapproach is unable to handle the user query.

\begin{myalgorithm}
    \renewcommand{\txtcode}[1]{\defaultalgofontsize \texttt{#1}}

    \caption{
    Runtime recovery}
    \label{algo:runtime-recovery}
    
    \Fn{RuntimeRecovery$(\mathrm{FailedFnCalls}~\mathcal{F}_\text{fail},~\mathrm{FRG}~G=<K,R_1,R_2>,~\mathrm{ScheduledCalls}~S,~\mathrm{CompletedCalls}~C,~\mathrm{ToolSet},~\mathrm{MaxAttempts}~A)$}{

        $A \gets 0$, $S \gets S \cap C$ \Comment{Remove unexecuted calls} \;
    
        \While{$\mathcal{F}_\text{fail} \neq \varnothing \land A < \mathrm{MaxAttempts}$}{
            $A \gets A + 1$ \;
            $N_\text{all} \gets \varnothing$ \Comment{Recovery points} \;
            
            \For{$\kappa_\text{fail} \in \mathcal{F}_\text{fail}$}{
                \If{$\callfn{reexecute}{\kappa_\text{fail}} == \textsf{success}$}{
                    $C \gets C \cup \{\kappa_\text{fail}\}$ \; 
                    $S \gets S \setminus \{\kappa_\text{fail}\}$ \Comment{Transient fault recovered} 
                }
                \Else{
                    $\mathcal{P} \gets \{\kappa' \in K \mid \langle \kappa', \kappa_\text{fail} \rangle \in R_2\}$ \;
                    \If{$\mathcal{P} == \varnothing$}{ $N_\text{all} \gets N_\text{all} \cup \{\kappa_\text{fail}\}$ }
                    \Else{
                        $r^* \gets \min\{\mathrm{rank}(\kappa') \mid \kappa' \in \mathcal{P}\}$ \Comment{Closest parent(s)} \;
                        $N_\text{all} \gets N_\text{all} \cup \{\kappa' \in \mathcal{P} \mid \mathrm{rank}(\kappa') = r^*\}$  
                    }
                }
            }

            \For{$\kappa_\text{rec} \in N_\text{all}$}{
                $\kappa_\text{new} \gets \callfn{repaircall}{\kappa_\text{rec}, \callfn{collectctx}{\kappa_\text{rec}}, \mathrm{ToolSet}}$ \Comment{Repair can modify arguments or replace the function/tool} \;
                \If{$\kappa_\text{new} \neq \textsf{null}$}{ 
                    $\callfn{replacenode}{\mathrm{G}, \kappa_\text{rec}, \kappa_\text{new}}$ \Comment{Updated node will be re-executed from this recovery point} \;
                    $\callfn{reschedule}{S, \kappa_\text{new}}$ \;
                    
                }
            }

            $(\mathcal{C}_\text{new}, \mathcal{F}_\text{fail}) \gets \callfn{execute}{N_\text{all}}$ \Comment{Execute all recovery points and their descendants; collect newly completed calls and new failures}  \;
            $C \gets C \cup \mathcal{C}_\text{new}$ \; 
            $S \gets (S \setminus \mathcal{C}_\text{new}) \cup \mathcal{F}_\text{fail}$ 
        }

        \Return \textsf{success} \text{ if } $\mathcal{F}_\text{fail} == \varnothing$ \text{ else } \textsf{failure} \;
    }
\end{myalgorithm}

\smalltitle{Runtime Recovery}
To avoid the significant overhead of restarting the entire orchestration from scratch, \ourapproach implements a fine-grained runtime recovery mechanism. We have observed that function call failures in LLM-driven agents typically stem from two root causes:
\begin{itemize}
    \item \textit{\textbf{Self-contained Fault}}: The function call itself is faulty (e.g., the target tool is temporarily unavailable, the function contains an internal logic error, or its parameters are malformed).
    \item \textbf{\textit{Data-Dependent Fault}}: The data provided by its parent call(s) is insufficient or incorrect, causing the current call to fail (e.g., a search call did not retrieve enough information for a subsequent math call to generate a valid expression).
\end{itemize}
\ourapproach's runtime recovery strategy, formalized in Algorithm \ref{algo:runtime-recovery}, is designed to diagnose the most likely cause and attempt a repair with minimal re-execution. The core idea is to identify a set of recovery points in the FRG.  \ourapproach first isolates errors by clearing all scheduled but unexecuted calls (Line 2), preventing propagation of irrelevant failures. 
For each failed call $\kappa_\text{fail}$, \ourapproach attempts a simple re-execution with the same arguments (Line 7) to detect self-contained faults, such as transient tool unavailability or malformed parameters; success marks the failure as recovered and removes it from the scheduled set (Line 8-9). 
If re-execution fails, the failure is treated as data-dependent, and \ourapproach traces back through the FRG to identify recovery points. Specifically, it selects the closest parent nodes $\kappa' \in K$ with minimal rank among those that have a data dependency on $\kappa_\text{fail}$ (Lines 14–16), prioritizing nodes that are nearest in the data flow. 
For each recovery point $\kappa_\text{rec} \in N_\text{all}$, the LLM generates a repaired call $\kappa_\text{new}$ (Lines 19–21), which may involve modifying arguments (e.g., search parameters) or replacing the function with an alternative from the user-defined toolset.
The repaired call then replaces the original in the FRG (Line 20), and execution is rescheduled from the recovery points onward (Line 22).
All recovery points and their descendants are executed, and newly completed calls and failures are updated accordingly (Lines 23-24).
This process iterates up to a predefined maximum number of attempts, after which the system reports success if all failures are resolved or failure otherwise (Line 25). 
\ourapproach provides a structured mechanism for fault localization, targeted repair, and selective re-execution.

\smalltitle{Discussion}
\ourapproach conservatively assumes that the root causes of errors occur in close proximity.
The current recovery mechanism is a foundational step towards more intelligent fault tolerance in LLM orchestration.
Nevertheless, there is a performance-accuracy trade-off when determining the depth of tracing in \ourgraphshort.
\ourapproach opts to trace back only to the closest parents, as this sufficed in our experiments.
For more complex user queries, determining the appropriate depth remains a challenge; we plan to address this in future work.
Additionally, the LLM may replace an inout function with a compute function (or vice versa) during runtime recovery, which could alter \ourgraphshort.
However, we did not observe such instances in our evaluation, and the current implementation of \ourapproach prevents this from occurring.
Moreover, \ourapproach does not support dynamic re-planning~\cite{kim2024an} as it is unable to obtain the correct answer for a practical user query.

\section{Experimental Evaluation}\label{sec:evaluation}

We structured our evaluation for \ourapproach around the following research questions:

\begin{enumerate}[leftmargin=2.5em, label=\textbf{RQ\arabic*}]
    \item How does \ourapproach compare to state-of-the-art techniques when orchestrating inout functions?
    Can it achieve comparable performance?
    Since inout functions are the most frequently called in today's LLMs, this research question seeks to determine whether \ourapproach compromises performance on inout functions to support compute functions.
    \conglidrawbacks{We missed WebShop}
    \conglidrawbacks{Also, perhaps looking for some benchmarks used by Open LLM Leaderboards, LMSys, etc.}

    \item When handling user queries with intensive compute function calls, how does \ourapproach fare?
    Can it deliver substantial efficiency improvements without compromising accuracy?
    The primary goal of \ourapproach is to support calling functions with heavy computation; this research question examines that objective.

    \item Does \ourapproach's recovery mechanism effectively function in RQ1 and RQ2?
    Can it address real-world errors encountered while handling user queries?
    On average, how many recovery attempts are needed to address a raised error?

    \item How does the performance of \ourapproach scale as more processors are allocated?

    \item Can \ourapproach be effectively utilized in real-world scenarios?
    \conglidrawbacks{Our case studies are not real-world; SWE-agent}
\end{enumerate}

\subsection{Experimental Setup}

\smalltitle{Baselines}
We selected ReAct~\cite{yao2022react}, Parallel Function Calling (ParallelFC)~\cite{openaipfcandex1}, and \llmcompiler~\cite{kim2024an} as baselines%
\footnote{We did not include \llmtoolcompiler in our evaluation as it is not yet open-source by the time of writing this paper.}
(\Cref{tab:comp-sota}),
and evaluated them on GPT-3.5-Turbo, 
GPT-4, and GPT-4o
.
\conglidrawbacks{Include LLM-Tool Compiler and Llama3-70b.}
To ensure a fair comparison, we reused---yet refined  to the best of our efforts---\llmcompiler's prompt and in-context examples to fit \ourapproach when translating user queries into function call sequences.
We also adhered to \llmcompiler's experimental setup when configuring ReAct and OpenAI's ParallelFC (or OpenAI PFC), including their prompts, in-context examples, and LLM parameters.
However, instead of sequentially calling OpenAI PFC-generated parallel function calls (as seen in their official examples and \llmcompiler's settings), we delegated the execution of function calls to Python's \txtcode{asyncio}---the same runtime used by \llmcompiler---to enable concurrency.


\conglidrawbacks{We missed/overlooked many important experiments that \llmcompiler has ever conducted.}

\smalltitle{Metrics}
We evaluated the techniques based on their \emph{accuracy}, \emph{latency speedup}, and \emph{token costs}, following \llmcompiler.
Accuracy refers to the correctness of the answer obtained after addressing a user query using a specific technique.
Latency represents the end-to-end runtime (in seconds) for addressing a user query.
Latency speedup measures the efficiency improvement achieved by a technique compared to ReAct, the sequential orchestration technique.
Token costs are the collective input and output tokens consumed during the addressing of a user query.

\smalltitle{Other Configurations}
Our experiments were conducted on an Apple Macbook Pro with macOS 12.1, 32~GiB RAM, and a 10-core M1 Pro CPU.
All results were obtained by averaging three runs.
GPT-3.5-Turbo (1106) and some functions (e.g., \txtcode{wiki}) are deployed remotely and we interacted with them through APIs.
Therefore, the latency presented in our results may be influenced by internet conditions, but we ensured all the techniques were assessed in a same environment.

\begin{table*}[tb]

  \centering
  \footnotesize

  \renewcommand{\arraystretch}{.8}
  \setlength{\tabcolsep}{1.15em}

  \caption{
    Comparison results for different benchmarks.
    \emph{Costs} indicates the token costs (dollars/1,000tokens).
    For techniques with recovery/replanning mechanisms,
    \emph{\#Recoveries} denotes the average number of recoveries required to address an error, or ``$\infty$'' if the errors cannot be addressed out of two attempts;
    \emph{\#Errors} ($x / y$) shows the number of successfully addressed errors ($x$) out of all raised errors ($y$).
    Note that RQ1 uses natural errors for recovery and RQ2 uses artificially constructed errors via fault injection.
  }
  \label{tab:eval_comp_res}

\begin{tabular}{cccccccccc}
\midrule
\textbf{Benchmark}                        & \textit{Model }                                                           & \textbf{Technique}                   &  & \textit{Accuracy} & \textit{Latency} & \textit{Speedup}                                                                               & \textit{Costs}   & \textit{\#Errors} & \textit{\#Recoveries} \\ \midrule
\multicolumn{10}{c}{\cellcolor{gray!15}\color{black!70} RQ1: I/O-Intensive Evaluation}
\\ \addlinespace[2pt]
\multirow{8}{*}{\textbf{HotpotQA}}                 & \multirow{4}{*}{GPT-3.5-Turbo}                                            & ReAct                       &  & 62.13\%  & 7.61s   & 1.00$\times$                                                                          & 5.00\$  & ---      & ---          \\
                                          &                                                                           & OpenAI PFC                  &  & 62.00\%  & 5.39s   & 1.41$\times$                                                                          & 1.57\$  & ---      & ---          \\
                                          &                                                                           & \llmcompiler &  & 62.00\%  & 5.02s   & 1.52$\times$                                                                          & 1.47\$  & 0 / 3    & $\infty$     \\
                                          &                                                                           & \ourapproach &  & 62.04\%  & 4.80s   & \textcolor{ColorTableGood}{\textbf{1.59$\times$}} & 1.401\$ & 3 / 3    & 1            \\ \addlinespace[2pt] \cline{2-10} \addlinespace[2pt]
                                          & \multirow{4}{*}{{GPT-4o}}                                                   & ReAct                       &  & 67.88\%  & 10.16s  & 1.00$\times$                                                                          & 17.44\$ & ---      & ---          \\
                                          &                                                                           & OpenAI PFC                  &  & 67.79\%  & 7.84s   & 1.30$\times$                                                                          & 5.42\$  & ---      & ---          \\
                                          &                                                                           & \llmcompiler &  & 67.83\%  & 7.48s   & 1.36$\times$                                                                          & 5.10\$  & 0 / 3    & $\infty$     \\
                                          &                                                                           & \ourapproach &  & 67.85\%  & 7.36s   & \textcolor{ColorTableGood}{\textbf{1.38$\times$}} & 4.93\$  & 3 / 3    & 1            \\ \midrule
\multirow{8}{*}{\textbf{MovieRec}}                 & \multirow{4}{*}{GPT-3.5-Turbo}                                          & ReAct                       &  & 77.20\%  & 25.20s  & 1.00$\times$                                                                          & 20.46\$ & ---      & ---          \\
                                          &                                                                           & OpenAI PFC                  &  & 77.40\%  & 10.07s  & 2.50$\times$                                                                          & 3.15\$  & ---      & ---          \\
                                          &                                                                           & \llmcompiler &  & 77.60\%  & 9.79s   & 2.57$\times$                                                                          & 3.04\$  & 0 / 3    & $\infty$     \\
                                          &                                                                           & \ourapproach &  & 77.40\%  & 8.27s   & \textcolor{ColorTableGood}{\textbf{3.04$\times$}} & 3.10\$  & 3 / 3    & 1            \\ \addlinespace[2pt] \cline{2-10} \addlinespace[2pt] 
                                          & \multirow{4}{*}{{GPT-4o}}                                                   & ReAct                       &  & 81.60\%  & 29.63s  & 1.00$\times$                                                                          & 55.24\$ & ---      & ---          \\
                                          &                                                                           & OpenAI PFC                  &  & 80.02\%  & 13.47s  & 2.20$\times$                                                                          & 8.51\$  & ---      & ---          \\
                                          &                                                                           & \llmcompiler &  & 81.60\%  & 12.99s  & 2.28$\times$                                                                          & 8.11\$  & 0 / 3    & $\infty$     \\
                                          &                                                                           & \ourapproach &  & 81.60\%  & 12.08s  & \textcolor{ColorTableGood}{\textbf{2.45$\times$}} & 8.18\$  & 3 / 3    & 1            \\ \midrule
\multirow{10}{*}{\textbf{ParallelQA}}               & \multirow{2}{*}{GPT-3.5-Turbo (x)}                                                     \\                                                                                                                                                        \\ \addlinespace[2pt] \cline{2-10} \addlinespace[2pt]
                                          & \multirow{4}{*}{{GPT-4o}}                                                   & ReAct                       &  & 76.87\%  & 36.25s  & 1.00$\times$                                                                          & 184\$   & ---      & ---          \\
                                          &                                                                           & OpenAI PFC                  &  & 74.50\%  & 23.00s  & 1.58$\times$                                                                          & 43\$    & ---      & ---          \\
                                          &                                                                           & \llmcompiler &  & 76.91\%  & 21.74s  & 1.67$\times$                                                                          & 46\$    & 0 / 9    & $\infty$     \\
                                          &                                                                           & \ourapproach &  & 76.99\%  & 18.39s  & \textcolor{ColorTableGood}{\textbf{1.97$\times$}} & 50\$    & 9 / 9    & 1            \\ \addlinespace[2pt] \cline{2-10} \addlinespace[2pt]
                                          & \multirow{4}{*}{GPT-4}                                                    & ReAct                       &  & 89.30\%  & 40.52s  & 1.00$\times$                                                                          & 480\$   & ---      & ---          \\
                                          &                                                                           & OpenAI PFC                  &  & 88.50\%  & 25.97s  & 1.56$\times$                                                                          & 121\$   & ---      & ---          \\
                                          &                                                                           & \llmcompiler &  & 88.90\%  & 25.44s  & 1.59$\times$                                                                          & 103\$   & 0 / 9    & $\infty$     \\
                                          &                                                                           & \ourapproach &  & 89.44\%  & 21.15s  & \textcolor{ColorTableGood}{\textbf{1.92$\times$}} & 118\$   & 9 / 9    & 1            \\ \midrule
\multicolumn{10}{c}{\cellcolor{gray!15}\color{black!70} RQ2: Compute-Intensive Evaluation}                                                                                                                                                                     \\ \addlinespace[2pt]
\multirow{8}{*}{\textbf{KITTI$^{\ast}$}}           & \multirow{4}{*}{\begin{tabular}[c]{@{}c@{}}GPT-3.5-Turbo\end{tabular}} & ReAct                       &  & 46.88\%  & 303.45s & 1.00$\times$                                                                          & 15.60\$ & ---      & ---          \\
                                          &                                                                           & OpenAI PFC                  &  & 46.87\%  & 277.67s & 1.09$\times$                                                                          & 2.51\$  & ---      & ---          \\
                                          &                                                                           & \llmcompiler &  & 46.87\%  & 278.68s & 1.09$\times$                                                                          & 2.40\$  & 0 / 0    & 0            \\
                                          &                                                                           & \ourapproach &  & 46.87\%  & 143.19s & \textcolor{ColorTableGood}{\textbf{2.12$\times$}} & 2.20\$  & 10 / 10    & 1            \\ \addlinespace[2pt] \cline{2-10} \addlinespace[2pt]
                                          & \multirow{4}{*}{{GPT-4o}}                                                   & ReAct                       &  & 46.88\%  & 308.30s & 1.00$\times$                                                                          & 39.78\$ & ---      & ---          \\
                                          &                                                                           & OpenAI PFC                  &  & 46.88\%  & 285.89s & 1.09$\times$                                                                          & 6.53\$  & ---      & ---          \\
                                          &                                                                           & \llmcompiler &  & 46.88\%  & 285.46s & 1.08$\times$                                                                          & 5.76\$  & 0 / 0    & 0            \\
                                          &                                                                           & \ourapproach &  & 46.88\%  & 149.43s & \textcolor{ColorTableGood}{\textbf{2.06$\times$}} & 5.39\$   & 10 / 10    & 1            \\ \midrule
\multirow{8}{*}{\textbf{AGNews$^{\ast}$}} & \multirow{4}{*}{\begin{tabular}[c]{@{}c@{}}GPT-3.5-Turbo\end{tabular}} & ReAct                       &  & 100.00\% & 58.72s  & 1.00$\times$                                                                          & 4.25\$  & ---      & ---          \\
                                          &                                                                           & OpenAI PFC                  &  & 100.00\% & 39.53s  & 1.49$\times$                                                                          & 2.33\$  & ---      & ---          \\
                                          &                                                                           & \llmcompiler &  & 100.00\% & 38.82s  & 1.51$\times$                                                                          & 2.20\$  & 0 / 0    & 0            \\
                                          &                                                                           & \ourapproach &  & 100.00\% & 29.44s  & \textcolor{ColorTableGood}{\textbf{1.99$\times$}} & 1.53\$   & 10 / 10    & 1            \\ \addlinespace[2pt] \cline{2-10} \addlinespace[2pt]
                                          & \multirow{4}{*}{{GPT-4o}}                                                   & ReAct                       &  & 100.00\% & 61.18s  & 1.00$\times$                                                                          & 10.69\$ & ---      & ---          \\
                                          &                                                                           & OpenAI PFC                  &  & 100.00\% & 43.11s  & 1.42$\times$                                                                          & 5.70\$  & ---      & ---          \\
                                          &                                                                           & \llmcompiler &  & 100.00\% & 42.29s  & 1.45$\times$                                                                          & 5.56\$  & 0 / 0    & 0            \\
                                          &                                                                           & \ourapproach &  & 100.00\% & 32.97s  & \textcolor{ColorTableGood}{\textbf{1.86$\times$}} & 3.76\$   & 10 / 10    & 1            \\ \midrule
\end{tabular}
\end{table*}

\subsection{RQ1. Effectiveness on I/O-Intensive Tasks}\label{ssec:evaluation_io}

Similar to \llmcompiler, we chose HotpotQA~\cite{yang2018hotpotqa}, Movie Recommendation (or MovieRec)~\cite{srivastava2022beyond}, and ParallelQA~\cite{kim2024an}
\conglidrawbacks{Add WebShop} as our benchmark datasets.
User queries of these datasets rely more on inout functions (\txtcode{wiki}, \txtcode{self}) than compute functions (\txtcode{math}, \txtcode{read}, \txtcode{write}), but are with different parallelism complexity~\conglidrawbacks{It would be better if we can measure the complexity of the FRG using max\_rank and max\_num\_funcs\_per\_rank}.
\begin{itemize}
    \item HotpotQA comprises 1,500 user queries designed to compare two distinct entities based on specific aspects, for example, ``Are both Duke Energy and Affiliated Managers Group based in Massachusetts?''.
    The dataset represents the lowest complexity level, typically involving two concurrent \txtcode{search}es and a final \txtcode{self} for summarization.

    \item MovieRec, derived from Google's open-source BIG-bench project, consists of 500 user queries.
    These queries task to select the most similar movie from a group of four movies in comparison to another group of four movies.
    This dataset demonstrates moderate complexity, typically involving eight concurrent \txtcode{search}es and a final \txtcode{self} for summarization.

    \item ParallelQA is specially crafted by \llmcompiler, containing 113 user queries.
    The \ourgraphshort constructed to address these queries typically follows three different, much more complex patterns, each involving 5--8 function calls and an average depth (i.e., the maximum rank) of four.

    \conglidrawbacks{WebShop}
\end{itemize}
We enforced two processors for HotpotQA and eight processors for other datasets.
\conglidrawbacks{and 3 processors for ParallelQA to best fit \llmcompiler}
\conglidrawbacks{Why do we enfores 2 and 3; it's too ad-hoc?}

\Cref{tab:eval_comp_res}'s ``RQ1: I/O-Intensive Evaluation'' section showcases the results.
Overall, \ourapproach achieved the highest latency speedup without compromising accuracy on all benchmark datasets, however, the average speedup of OpenAI PFC, \llmcompiler, and \ourapproach are all comparable with around 1.82$\times$, 1.89$\times$, and 2.18$\times$, respectively.
This situation is because these datasets rely more on inout function calls which are quickly suspended waiting for the I/O to finish, during when their processors switch to another function call for all techniques except ReAct.
Among the three datasets, all techniques brought more speedups on the MovieRec benchmark possibly due to the higher number of concurrent function calls at a time for each user query.
In this benchmark, \ourapproach accelerated the process by an average of 3.04$\times$.
It is noteworthy that, compared to the state-of-the-art \llmcompiler, the speedup of \ourapproach increased with the complexity of the dependencies (FRGs), even in such IO-intensive circumstances.
Specifically, \ourapproach achieves the highest speedup ($\sim$1.21$\times$) on ParallelQA--the most complex benchmark---with 5--8 function calls and an average FRG depth of four---evaluated in this research question.
In terms of token costs, \ourapproach achieved a reduction of up to 3.57$\times$, 6.60$\times$ and 4.07$\times$ for HotpotQA, MovieRec and ParallelQA.
This likely contributes to the query translation step in \llmcompiler and \ourapproach, which removes the need to iteratively interacting with LLMs and to generating a thought at each single step as in ReAct.
OpenAI PFC is also the former case.

An interesting observation is that GPT-4o exhibits a smaller relative speedup than GPT-3.5-Turbo. This is due to its less optimized native parallelism and the fact that, for larger and more advanced models, the sequential stages of the workflow—such as query translation, relation discovery, and summarization—take longer because of increased internal reasoning complexity. This is consistent with Amdahl’s Law~\cite{10.1145/1465482.1465560}. These findings suggest that \ourapproach provides the most significant benefits in scenarios where sequential costs are lower and execution dominates, such as with faster models or resource-constrained settings, highlighting its role as a model-agnostic optimization layer.

\boxedtext{\textbf{Answer to RQ1:} 
    \ourapproach demonstrates superior performance, particularly in latency, to state-of-the-art techniques when orchestrating inout function calls, while also maintaining competitive token costs and similar or improved accuracy.
}

\subsection{RQ2. Effectiveness on Compute-Intensive Tasks}\label{ssec:evaluation_compute}

As we did not find suitable datasets for this new problem, we specially adapted the widely used KITTI dataset~\cite{Geiger2012CVPR} in autonomous driving 
and AGNews dataset~\cite{AGNews} in text classification 
as our benchmark dataset.
We named them KITTI$^\ast$ and AGNews$^\ast$.

\begin{itemize}
    \item KITTI$^{\ast}$.
    We sampled 1,600 images and created 200 user queries accordingly to simulate real-world differential testing of two autonomous driving systems.
    For each query, we tasked it to calculate the difference of the steering angles of two groups of scenes, with each group containing four scenes from one system.
    This involves eight concurrent, compute-intensive \txtcode{stereorcnn}es for steering angle detection, followed by two concurrent \txtcode{self}s for averaging the angles per group and a final \txtcode{self} to output their difference.
    \item AGNews$^\ast$.
    The dataset comprises 120,000 samples from over 2,000 news sources.
    We created 120 user queries, each utilizing 1,000 samples to simulate a real-world data mining and analysis task, specifically for t-distributed Stochastic Neighbor Embedding (t-SNE) and Latent Dirichlet Allocation (LDA).
    Each query consists of five function calls:
    a \txtcode{self} to generate guidelines for the task,
    a \txtcode{read} to retrieve 1,000 news from the dataset,
    two concurrent compute-intensive calls \txtcode{t-SNE} and \txtcode{LDA} to perform data mining and visualization, 
    and a final \txtcode{write} to export the results.
\end{itemize}
We enforced four processors for KITTI$^{\ast}$ and two processors for AGNews$^{\ast}$.


\begin{figure}
    \centering
    \includegraphics[width=\linewidth]{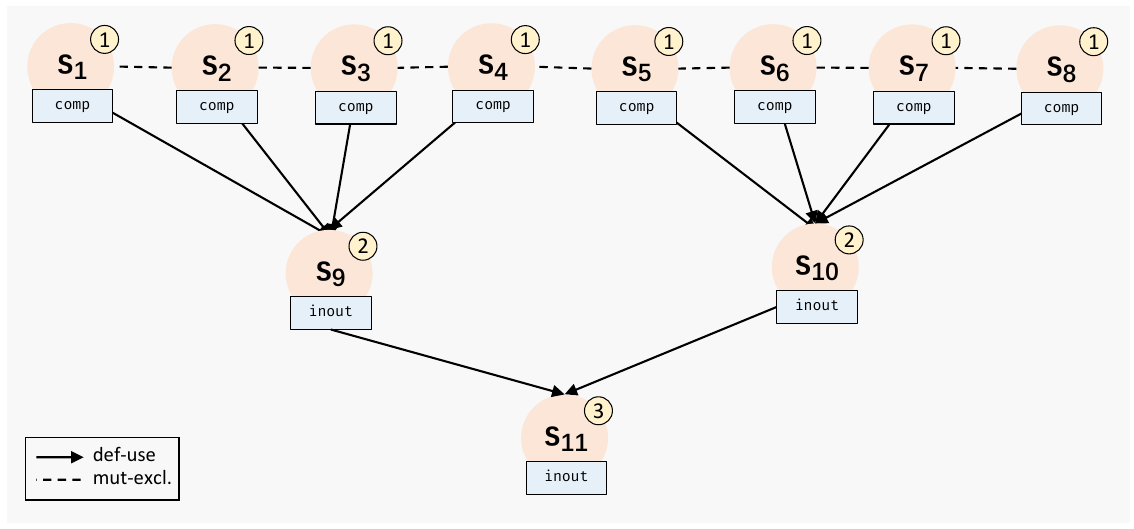}
    \caption{
        \ourgraphshort for The Example in RQ2.
    }
    \label{fig:eval_rq2_ex}
\end{figure}

The ``RQ2: Compute-Intensive Evaluation'' section in \Cref{tab:eval_comp_res} displays the results.
In contrast to the previous experiment, the speedup introduced by \ourapproach were significantly higher (at least 1.32$\times$) even than \llmcompiler.
\ourapproach's speedup over \llmcompiler reached even 1.94$\times$ on the KITTI$^{\ast}$ benchmark.
We attribute this to the discovered control relations and the separation of scheduling and execution on a resource-constrained machine (four available processors with eight concurrent compute function calls, or two available processors with five function calls).
In \llmcompiler, all concurrent, computing function calls are executed simultaneously immediately after they are submitted, resulting in a $>$2$\times$ overburden of processors.
Actually, the \txtcode{asyncio}'s job scheduler schedules them sequentially in a single thread as they block the processor.
In this case, the kernel's thread scheduler again schedules the single thread together with other threads periodically.
In contrast, the four additional function calls are not seen by the kernel's thread scheduler until our coordinator coordinates them to a specific processor.
As for AGNews$^{\ast}$, the speedup of \ourapproach was less significant than that on KITTI$^{\ast}$ as the total number of concurrent compute calls is 2, aligning with the number of processors that we have allocated to it.
As for token costs, \ourapproach achieved a reduction of around 7.09$\times$ on KITTI$^{\ast}$ benchmark and 2.78$\times$ on AGNews$^{\ast}$ benchmark, similar to OpenAI PFC and \llmcompiler.

\smalltitle{KITTI$^\ast$ Example}
The following displays a (simplified) example of the KITTI$^{\ast}$ benchmark:

\begin{userprompt}
We collect stereo images for both systems A and B, named from 000001.png to 000008.png, with odd-numbered images for system A and even-numbered images for system B. Then, process all images using the 3D object detection algorithm stereorcnn.py to calculate the steering angles for each system across different 4 perspectives. Compute the average steering angles of the two systems using a chatbot. Finally, calculate the difference in average steering angles between systems A and B to quantify the difference in their stability.
\end{userprompt}

Upon receiving the query, \ourapproach constructed its \ourgraphshort as in \Cref{fig:eval_rq2_ex}.
The graph involved two sets of concurrent steering angle detection tasks, each set employing the compute function \txtcode{stereorcnn} four times ($s_1$--$s_4$ and $s_5$--$s_8$), followed by a inout function call \txtcode{self} ($s_9$ and $s_{10}$) to average all detected angles per group.
The final difference of the two systems was output with an additional, inout function call \txtcode{self} ($s_{11}$).
When coordinating them to four processors, $s_1$--$s_4$ were executed initially on four independent processors.
After their completion, $s_5$--$s_8$ were executed on the same 4 independent processors.
Subsequently, $s_9$ and $s_{10}$ were executed on an arbitrary processor as they were inout calls.
Lastly, $s_{11}$ was executed on an arbitrary processor.

\smalltitle{AGNews$^\ast$ Example}
Below is a (simplified) example for the AGNews$^{\ast}$ benchmark:
\begin{userprompt}
I want to visualize the collected news data. I will read dataset details from readme.txt. Then use the tsne.py script to visualize the distribution of texts in the dataset, and use the lda.py script to visualize the topic modeling of the dataset. Meanwhile, I need a brief sample procedure for t-SNE and LDA analysis on NLP tasks from the chatbot. Finally, all outputs should be saved in a folder called /result.
\end{userprompt}

Despite the small number of function calls (only five), the dependencies within each user query are non-trivial.
Indeed, the FRG and the overall scheduling procedure are consistent with \Cref{fig:overview}.

\boxedtext{\textbf{Answer to RQ2:}
    Compared to state-of-the-art techniques, \ourapproach demonstrated a significantly superior speedup in latency for compute-intensive function orchestrations without compromising the accuracy.
    The token costs of \ourapproach is comparable to them.
}

\begin{table*}[tb]
    \centering
    \footnotesize

    \renewcommand{\arraystretch}{.8}
    \setlength{\tabcolsep}{1.2em}

    \caption{
        Statistics of latency speedups with allocated processors on the crafted KITTI benchmark.
        Columns ``\emph{Spd.}'' represent speedups.
        The latency speedups are calculated between a respective technique and ReAct.
    }
    \label{tab:eval_perf_trend}
    
    \begin{tabular}{lcccccccc}
    \toprule
    \multirow{2}{*}{\bf Technique} &
    \multicolumn{2}{c}{1 Processor} &
    \multicolumn{2}{c}{2 Processors} &
    \multicolumn{2}{c}{3 Processors} &
    \multicolumn{2}{c}{4 Processors} \\ 
    \cmidrule(lr){2-3}
    \cmidrule(lr){4-5}
    \cmidrule(lr){6-7}
    \cmidrule(lr){8-9}
    &
    \emph{Latency} & \emph{Spd.} &
    \emph{Latency} & \emph{Spd.} &
    \emph{Latency} & \emph{Spd.} &
    \emph{Latency} & \emph{Spd.} \\
    \midrule
    ReAct           & 328.08s   & 1.00$\times$   & 316.66s 
    & 1.04$\times$  & 312.12s   & 1.05$\times$   & 303.45s 
    & 1.08$\times$ \\
    OpenAI PFC      & 290.82s   & 1.13$\times$   & 284.68s 
    & 1.15$\times$  & 282.63s   & 1.16$\times$   & 277.67s 
    & 1.18$\times$ \\
    \llmcompiler    & 286.47s   & 1.15$\times$   & 284.21s 
    & 1.15$\times$  & 280.85s   & 1.17$\times$   & 278.68s 
    & 1.18$\times$ \\
    \ourapproach    & 280.36s   & \textcolor{ColorTableGood}{\bf 1.17$\times$}  & 218.52s & \textcolor{ColorTableGood}{\bf 1.50$\times$}  & 208.19s & \textcolor{ColorTableGood}{\bf 1.58$\times$}  & 143.19s & \textcolor{ColorTableGood}{\bf 2.29$\times$} \\ 
    \midrule
    \multirow{2}{*}{\bf Technique} &
    \multicolumn{2}{c}{5 Processors} &
    \multicolumn{2}{c}{6 Processors} &
    \multicolumn{2}{c}{7 Processors} &
    \multicolumn{2}{c}{8 Processors} \\ 
    \cmidrule(lr){2-3}
    \cmidrule(lr){4-5}
    \cmidrule(lr){6-7}
    \cmidrule(lr){8-9}
     &
    \emph{Latency}    & \emph{Spd.}  &
    \emph{Latency}    & \emph{Spd.}  &
    \emph{Latency}    & \emph{Spd.}  &
    \emph{Latency}    & \emph{Spd.}  \\  
    \midrule
    ReAct        & 290.82s  & 1.13$\times$  & 286.47s  &
    1.15$\times$ & 284.85s  & 1.15$\times$  & 279.24s  &
    1.17$\times$ \\
    OpenAI PFC   & 275.54s  & 1.19$\times$  & 270.24s  &
    1.21$\times$ & 265.24s  & 1.24$\times$  & 264.47s  &
    1.25$\times$ \\
    \llmcompiler & 272.04s  & 1.21$\times$  & 269.81s  &
    1.22$\times$ & 267.11s  & 1.23$\times$  & 262.24s  &
    1.25$\times$ \\
    \ourapproach & 156.20s & \textcolor{ColorTableGood}{\bf 2.10$\times$} & 149.63s & \textcolor{ColorTableGood}{\bf 2.19$\times$} & 134.72s & \textcolor{ColorTableGood}{\bf 2.44$\times$} & 100.53s & \textcolor{ColorTableGood}{\bf 3.26$\times$} \\ 
    \bottomrule
    \end{tabular}
\end{table*}

\begin{figure*}[tb]
  \centering
  \includegraphics[width=.6\linewidth]{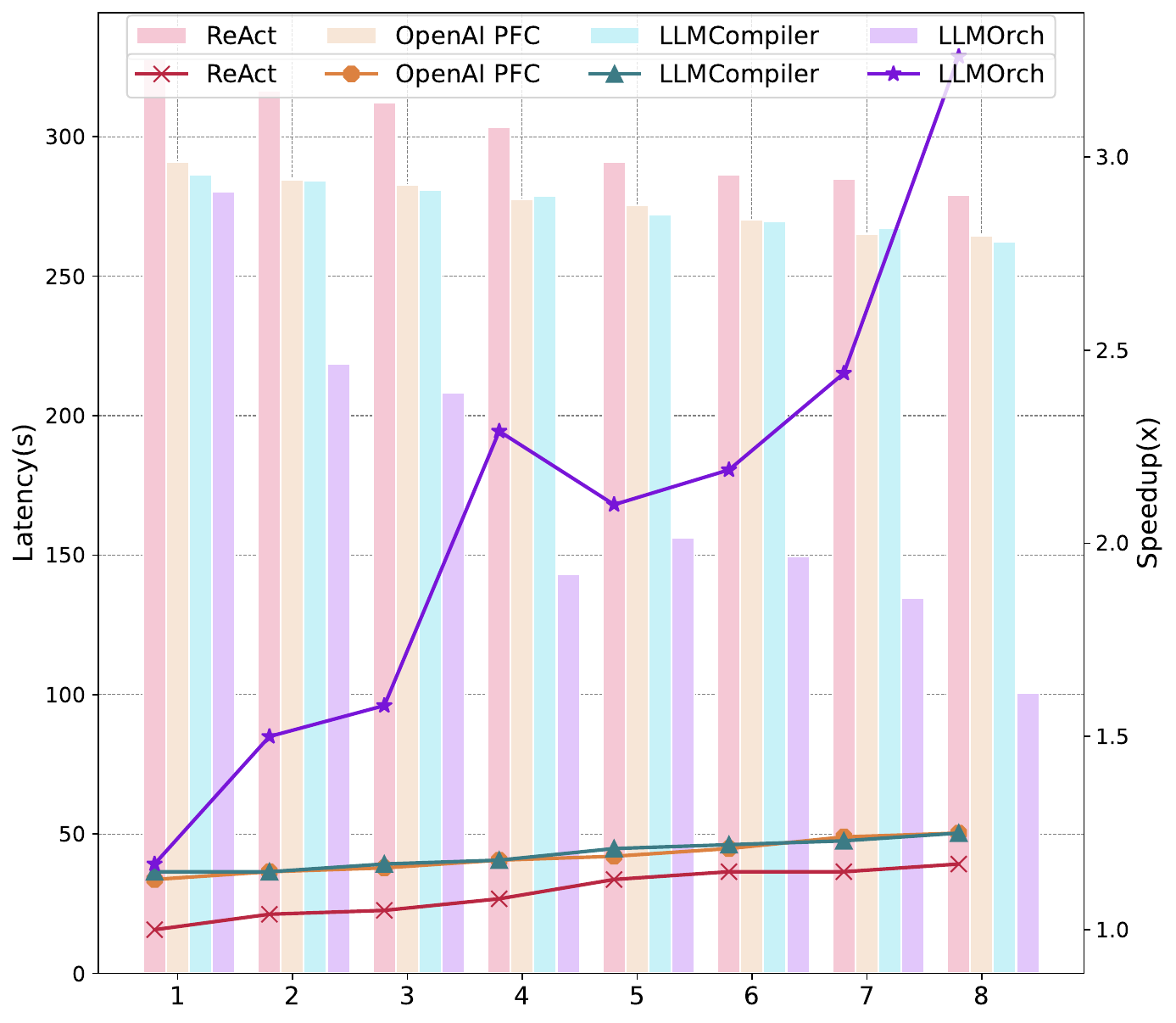} 
  \caption{Latency speedups with allocated processors on KITTI$^\ast$.}
  \label{fig:eval_perf_trend}
\end{figure*}

\subsection{RQ3. Errors and Recovery}\label{ssec:replan}

To evaluate whether \ourapproach's recovery mechanism effectively addresses errors, we recorded instances of errors and recoveries during previous experiments.
This investigation focused solely on \llmcompiler and \ourapproach, as ReAct and OpenAI's PFC do not support replanning or recovery.
We granted \llmcompiler and \ourapproach two attempts.

The last two columns (\emph{\#Errors} and \emph{\#Recoveries}) in \Cref{tab:eval_comp_res} summarize the results.
Overall, we observed three (out of 1,500), three (out of 500), and nine (out of 113) runtime errors in HotpotQA, MovieRec, and ParallelQA, respectively.
We did not find any compile-time errors, possibly contributing to the simple few-shot examples (rather than the complex grammar) we utilized for user query translation.
No errors were raised in KITTI$^{\ast}$ and AGNews$^{\ast}$, likely due to the straightforwardness of the user queries that we designed.
To systematically evaluate recovery robustness, we deliberately injected 20 Data-Dependent faults (10 per dataset) into KITTI$^{\ast}$ and AGNews$^{\ast}$ by providing compute functions with incomplete or malformed parameters. 
These results also match the instinct that complex queries, characterized by numerous function calls and stricter input/output formats, e.g., invoking the ``math'' function in ParallelQA, are more susceptible to exceptions.

For all the runtime errors, \ourapproach succeeded on the first attempt, while \llmcompiler failed both times.
After manual inspection, we realized that the failure of \llmcompiler was due to its (static) replanning strategy, which repeated the same operation without adjusting to runtime dynamic changes during execution (for example incorporating runtime feedback).

\smalltitle{Example: ParallelQA \#83}
Below is an example that demonstrates \ourapproach's recovery mechanism.
In this instance, all three baseline techniques (ReAct, ParallelFC, and \llmcompiler) failed to handle the user query.
ReAct and ParallelFC do not have replanning or recovery capabilities, while \llmcompiler attempted replanning but still failed.
In contrast, \ourapproach was able to resolve the runtime errors within a single recovery attempt and successfully obtained the correct answer (488.19).
The user query used in this example is as follows:

\begin{figure*}[tb]
    \centering
    \begin{minipage}[t]{0.48\textwidth}
        \input{figures/fig_recovery_llmcompiler}  
    \end{minipage}
    \hfill
    \begin{minipage}[t]{0.48\textwidth}
        \input{figures/fig_recovery_llmorch}
    \end{minipage}
\end{figure*}

\begin{userprompt}
If Texas and Florida were to merge and become one state, as well as California and Michigan, what would be the largest population density among these 2 new states and New Jersey? Answer in people / square km.
\end{userprompt}
All available functions are:
\begin{itemize}[leftmargin=1em]
    \item \txtcode{search(term,k=500)}:
    Search \txtcode{term} in Wikipedia and obtain the first \txtcode{k} words into a summary.
    \item \txtcode{math(prompt)}:
    Ask an LLM-based calculator for mathematics results by generating a rigorous mathematical expression that conforms to Python syntax based on the \txtcode{prompt}.
\end{itemize}
The function call sequences initially generated by \llmcompiler and \ourapproach are shown in \Cref{fig:recovery-example-llmcompiler} and \Cref{fig:recovery-example-llmorch}, respectively.

\emph{Errors}.
When handling the query, \ourapproach and \llmcompiler may throw errors at varied \txtcode{math} calls: \txtcode{s3}/\txtcode{5}, \txtcode{s4}/\txtcode{6}, \txtcode{s12}/\txtcode{14}, \txtcode{s5}/\txtcode{16} or \txtcode{s10}/\txtcode{17}.
The errors occur because the preceding \txtcode{search} (e.g., \txtcode{s2}/\txtcode{4}) did not retrieve enough information (500 words by default), causing the subsequent \txtcode{math} to lack valid data in generating a rigorous mathematical expression, finally leading to execution failures.

\emph{Recovery}.
As it was unable to obtain the correct answer for the user query, \llmcompiler's dynamic replanning capability is downgraded to static replanning, which requires LLMs to regenerate the overall function call sequence and reschedule all new function calls to run---similar to \ourapproach's compile-time recovery.
Despite this recovery, \llmcompiler still failed within two additional attempts even though with different function call sequences, raising the same errors at \txtcode{math}.
In contrast, \ourapproach captured the failed \txtcode{math} calls \txtcode{s3} and \txtcode{s4} with the exception message ``can't extract the population/area of Florida'' and started tracing back through \ourgraphshort.
In this example, the recovery call tracked was \lstcode{s2}.
Subsequently, the failed calls \txtcode{s3} and \txtcode{s4}, along with the exception message and all available functions, were sent to GPT-3.5-Turbo for repair of \txtcode{s2}.
GPT-3.5-Turbo successfully identified the error and updated the \txtcode{k} argument to \txtcode{1000}, resulting in a call that provided the information for Florida's population and area.
Finally, \ourapproach resumed scheduling from \lstcode{s2}.

\boxedtext{\textbf{Answer to RQ3:} 
    \ourapproach's recovery mechanism effectively functioned in addressing the runtime errors during RQ1 and RQ2 within one recovery attempts.
}

\subsection{RQ4. Performance on Different \#Processors}\label{ssec:performance_trend}

This experiment concerns how \ourapproach's performance scale in terms of the number of allocated processors.
In this experiment, we reused the KITTI$^{\ast}$ benchmark.
We accessed \ourapproach's performance against ReAct's and \llmcompiler's on processors ranging from one to eight--the maximum number of concurrent function calls allowed at a time in the benchmark.
In this research question, all speedup are calculated based on ReAct's latency on the one processor setting.

\Cref{tab:eval_perf_trend} presents the results with \Cref{fig:eval_perf_trend} plotting the trend of speedups.
We found that the speedup of \ourapproach scales nearly \emph{linearly} as the number of allocated processors increases continuously, except for a sudden latency reduction due to scheduling when allocating 4 processors.
Particularly, the speedup reached up to 3.26$\times$ when allocating eight processors.
In contrast, the state-of-the-art tool \llmcompiler achieved only up to 1.25$\times$ with speedups increasing seemingly logarithmically and OpenAI PFC was close to it.
They execute all concurrent function calls in parallel immediately after they are scheduled.
ReAct was hardly influenced by the number of processors as it is sequential.
It is worthwhile mentioning that when the number of allocated processors reached 4, \ourapproach significantly (nearly 2$\times$) outperformed the other two parallel techniques, thanks to our execution coordination, which alleviated the work burden of the underlying processors.
It should be noted that, when allocating 5 to 7 processors, the speedup of \ourapproach appears to be similar to that of four processors.
This is because the compute function calls ($s_1$--$s_8$ in \Cref{fig:eval_rq2_ex}) in each query of the KITTI$^{\ast}$ dataset are homogeneous, taking a comparable amount of time to complete.
As a result, one of the input function calls (e.g., $s_{10}$) must wait until all its predecessors finish.
This issue is alleviated with eight processors, as they tend to stop nearly simultaneously.

\boxedtext{\textbf{Answer to RQ4:}
    The speedup achieved by \ourapproach demonstrates a linear relationship with the number of allocated processors, while the accuracy is maintained within a marginal range.
}

\begin{figure}[tb]
\centering
\includegraphics[width=\linewidth]{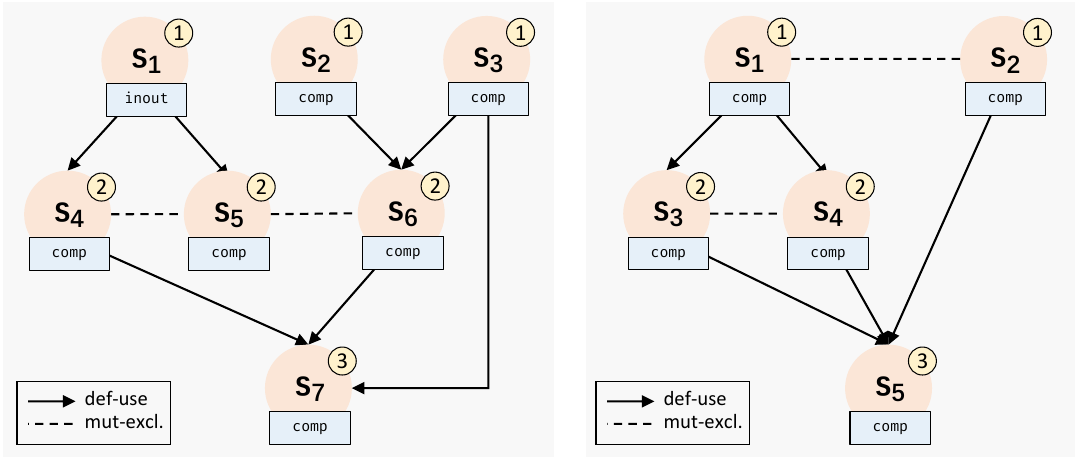}
\caption{
    The \ourgraphshort for case studies ``Purchase Intent Analysis'' (left) and ``End-to-End Encryption'' (right).
}
\label{fig:case_study_frgs}
\end{figure}

\subsection{RQ5. Real-World Case Studies}\label{ssec:case_studies}
Finally, we conducted two further real-world case studies to assess the practicability of \ourapproach, in addition to our illustrative example (\Cref{sec:illustrative_example}).

\smalltitle{Purchase Intent Analysis}
This is widely used in e-commerce and digital marketing.
It summarizes end users' purchase behaviors by analyzing their behavioral data, interests, preferences, and purchase history, playing a significant role in these industries.
In this case, we plan to treat LLMs as a recommendation system to analyze a user's purchasing behavior.
Specifically, based on the purchase history of a user, we requested LLMs to predict whether the user would be interested in purchasing Apple products.
We provide the following (simplified) prompts:
\begin{userprompt}
Treat yourself as a personalized user recommendation system to analyze users' purchase intentions towards some products. First, search Apple products via https://www.apple.com/. Use chatbot to generate a tutorial for analyzing user portraits. Then analyze the user's purchase history (history.txt) by the script analyze.py. Then use the content analysis script content.py and the user experience analysis script user.py I provided to analyze the product homepage. Let chatbot build a user portrait based on the historical analysis and the obtained tutorial. Finally, judge the user's purchasing behavior based on the product analysis, the user experience analysis, and user portrait.
\end{userprompt}
and available functions:
\begin{itemize}[leftmargin=1em]
    \item \lstcode{search(query)}:
    Search \lstcode{query} through search engines and obtain the results.
    \item \lstcode{chatbot(prompt)}:
    Ask an LLM for an answer or a completion towards \lstcode{prompt}.
    \item \lstcode{python(file,args)}:
    Execute the python code in \lstcode{file} with arguments \lstcode{args}.
\end{itemize}
As in our illustrative example, we deployed a local LLaMA3-8B as \lstcode{chatbot}, rending it being a compute function as \lstcode{python}.
\lstcode{search} is an inout function.
It should be noted that we intentionally streamlined the user query so that our LLMs can parse and generate a correct function call sequence for this complex scenario, as \ourapproach does not have a re-planning component.

\ourapproach obtained the following function call sequence.
\begin{funcseq}
s1: search(query="https://www.apple.com/")
s2: chatbot("Generate a tutorial for analyzing user portraits.")
s3: python(file="analyze.py", args=["history.txt"])
s4: python(file="content.py", args=[s1])
s5: python(file="user.py", args=[s1])
s6: chatbot(
    "Build a user portrait based on the following historical record analysis and user portrait
     tutorial: 
     ## Historical Record Analysis
     {s3} 
     ## User Portrait Tutorial
     {s2}"
)
s7: chatbot(
    "Judge the user's purchasing behavior based on the following product analysis, historical
     record analysis and user portrait:
     ## Product Analysis 
     {s4}
     ## Historical Record Analysis
     {s3}
     ## User Portrait
     {s6}"
)
\end{funcseq}
It can be observed that the user query is processed into seven function calls, including one inout function call (\lstcode{s1}) and six compute function calls (\lstcode{s2}--\lstcode{s7}).
The left sub-figure of \Cref{fig:case_study_frgs} presents the \ourgraphshort for them.

When running on three processors, \ourapproach was able to respond in around 77.10s, $\sim$1.71$\times$ (132.53s), $\sim$1.54$\times$ (119.00s), and $\sim$1.48$\times$ (114.62s) faster than ReAct, OpenAI PFC, and \llmcompiler, respectively, and all techniques output a similar, satisfactory response.
It is observed when handling queries involving relatively complex data and control relations, \llmcompiler exhibits slight inefficiencies in scheduling, whereas \ourapproach, achieves a certain improvement in processing speed.
However, we also find that it remains challenging for OpenAI PFC, \llmcompiler, and even \ourapproach to achieve an optimal acceleration even with increased computational resources when the topological relationships between function calls become complex.
We discussed this limitation and possible future works in \Cref{ssec:evaluation_discussion}.

\smalltitle{End-to-End Encryption} 
End-to-end encryption is commonly employed in practice, where a message sender encrypts a message and a message receiver decrypts it.
In this example, we aim to simulate a real-world scenario where an end user without programming skills seeks to utilize LLMs to (1) encrypt his chat (specifically the video and text record) using some provided encryption scripts and (2) obtain the encryption statistics towards them.

We provided \ourapproach with below (simplified) user query:
\begin{userprompt}
I want to do end-to-end encryption for privacy protection during a video call. I have a video file myvideo.mp4 and a chat text file chat.txt. I first need to generate a key for AES encryption and also obtain an example to evaluate the encryption effect (seems like the chatbot can do both). Then use the ASE encryption tool to encrypt the video and text. Finally, analyze the results of encryption by learning from the example and tell me.
\end{userprompt}
In addition to \lstcode{chatbot(prompt)}, we further offered \lstcode{aes(action,file,key)} which returns the statistics (e.g., time, ratio) of encrypting/decrypting (specified by \lstcode{action}) \lstcode{file} with \lstcode{key}.
Both functions are compute functions.

\ourapproach obtained the following function call sequence:
\begin{funcseq}
s1: chatbot("Generate a key for aes encryption.")
s2: chatbot("Give me an example to analyze the encryption effect.")
s3: aes(action="encrypt", file="myvideo.mp4", key=s1)
s4: aes(action="encrypt", file="chat.txt", key=s1)
s5: chatbot(
    "Learn from the following example on how to analyze the encryption effect: {s2}
     Then summarize the results of the following two encryptions: {s3} and {s4}."
)
\end{funcseq}
The \ourgraphshort is displayed in the right sub-figure of \Cref{fig:case_study_frgs}.

\ourapproach responded in approximately 53.82 seconds, while ReAct, OpenAI PFC, and \llmcompiler took approximately 109.11s ($\sim$2.03$\times$), 103.09s ($\sim$1.92$\times$), and 101.26s ($\sim$1.88$\times$), respectively, when running on two processors.
This notable efficiency improvement can be attributed to \ourapproach's coordination of the relationship between compute function calls and available processors.
Specifically, in this case where no input functions are given, the advantages of \llmcompiler and OpenAI PFC over the sequential method ReAct become minimal.
Furthermore, all techniques successfully encrypted both files.

\boxedtext{\textbf{Answer to RQ5:} 
    \ourapproach is applicable to addressing real-world scenarios, providing satisfactory answers as state-of-the-art techniques while bringing superior improved efficiency.
}

\subsection{{Ablation Study}}\label{ssec:abla}
To ensure that \ourapproach's performance gains are attributable to its orchestration algorithms rather than prompt engineering, we conducted a systematic ablation study to measure the variance induced by different few-shot examples.

\smalltitle{Experimental Setup} 
{\textit{\textbf{ 1) Dataset: }} We selected all five diverse datasets in the above experiments, i.e., HotpotQA, MovieRec, ParallelQA, KITTI*, and AGNews*. To balance statistical robustness with computational costs, we used randomly sampled subsets from each dataset:
300 queries (20\%) from HotpotQA (1,500 total), 100 queries (20\%) from MovieRec (500 total), and the full ParallelQA dataset (113 queries) due to its smaller size. Additionally, we used 40 queries (20\%) from KITTI* (200 total) and 24 queries (20\%) from AGNews* (120 total) due to the high execution time of each query in these benchmarks.
\textit{\textbf{2) Few-Shot Setting: }} We conducted five independent runs for each dataset, using GPT-3.5-Turbo and GPT-4 models. The number of few-shot examples was fixed for each dataset: 2 examples for HotpotQA, 1 for MovieRec, 5 for ParallelQA, and 1 for both KITTI* and AGNews*. The content of the few-shot examples varied randomly across runs, while all other components (e.g., prompt structure, hyperparameters) were kept constant.}

\smalltitle{Conclusion}
{The results, presented in Table \ref{tab:abla}, show minimal performance variability with the random selection of few-shot examples. Speedup differences between the best and worst runs were acceptable, with maximum variations of 0.03$\times$ for HotpotQA, 0.02$\times$ for MovieRec, and 0.03$\times$ for ParallelQA. For the custom datasets (KITTI* and AGNews*), the differences were even smaller (0.01$\times$). The planning success rate was consistently high across all datasets, with HotpotQA, MovieRec, KITTI*, and AGNews* achieving an average success rate of over 96\%. ParallelQA exhibited slightly higher variability (89\%), due to the dataset's increased query diversity and complexity. 
The choice of few-shot examples has a reasonable impact on the end-to-end latency of \ourapproach, with performance improvements remaining stable and reproducible across runs. These results confirm that the gains are driven by our novel runtime orchestration algorithms, rather than by the upfront prompt design or pre-defined few-shot examples.}

\begin{table}[b]
\caption{Ablation Study on Few-Shot Example.}
\label{tab:abla}
\resizebox{\columnwidth}{!}{%
\begin{tabular}{ccccccc}
\hline
\textbf{Dataset}    & \textbf{Model}   & \textbf{\#Size} & \textbf{\#Runs} & \textbf{\#Few-Shots} & \begin{tabular}[c]{@{}c@{}}\textbf{Speedup} \\ \textbf{Difference} \\ \textbf{(Max.)}\end{tabular} & \begin{tabular}[c]{@{}c@{}}\textbf{Planning} \\ \textbf{Success} \textbf{Rate} \\ \textbf{(Avg.)}\end{tabular} \\ \hline
HotpotQA   & GPT-3.5 & 300  & 5    & 2         & 0.03×                                                                   & 0.96                                                                       \\
MovieRec   & GPT-3.5 & 100  & 5    & 1         & 0.02×                                                                   & 0.99                                                                       \\
KITTI*     & GPT-3.5 & 40   & 5    & 1         & 0.01×                                                                   & 0.99                                                                       \\
AGNews*    & GPT-3.5 & 24   & 5    & 1         & 0.01×                                                                   & 0.99                                                                       \\
ParallelQA & GPT-4   & 113  & 5    & 5         & 0.03×                                                                   & 0.89                                                                       \\ \hline
\end{tabular}%
}
\end{table}

\subsection{Discussion}\label{ssec:evaluation_discussion}
In summary, ParallelFC, \llmcompiler, and \ourapproach were all successful in reducing the latency for solving user queries and meanwhile in maintaining the accuracy within a marginal range in our evaluation, when compared with ReAct.
As for \ourapproach, it demonstrated comparable speedups to the others when orchestrating inout function calls (\Cref{ssec:evaluation_io}), while significantly outperforming them for compute function calls (\Cref{ssec:evaluation_compute}).
It was also observed that the speedups achieved by \ourapproach had a linear relationship with the number of allocated processors (\Cref{ssec:performance_trend}).
When applying \ourapproach in real-world scenarios ``Purchase Intent Analysis'' and ``End-to-End Encryption'', it successfully addressed the given user queries with a correct answer (\Cref{ssec:case_studies}).
We believe that these results showcase the usefulness and practicability of \ourapproach in parallel function orchestration, specifically thanks to its discovery of data and control relations, as well as the separation of function call scheduling and execution.

\smalltitle{Limitations}
Despite the promising evaluation results \ourapproach has achieved, we realized that \ourapproach is primarily limited by its direct query translation, which translates a user query into a function call sequence without conditional (if/loop) structures.
While this is effective for common scenarios, such as those in our and \llmcompiler's evaluations, it falls short when the user query is complex and conditional, or if LLMs fail to produce a correct function call sequence even after our compile-time recovery.
Another limitation is that the current \ourapproach only supports an unchanged \ourgraphshort during runtime recovery.
For complex user queries that require numerous reasoning steps, where accuracy is more critical than performance, limiting \ourapproach to trace back to the closest recovery point and disallowing replacing inout functions with compute functions (and vice versa) may restrict \ourapproach's effectiveness.

\smalltitle{Future Work}
Beyond addressing the above limitations, there are several potential, exciting extensions for enhancing \ourapproach in the future:
(1) Modeling the side effects of functions.
In this work, we only consider the def-use (data) relations assuming that functions are pure.
However, it is likely in practice that a function call alters its used arguments or even a global resource, essentially re-defining them.
By modeling the side effects of functions, a widely used practice in compilers and language virtual machines, can help capture this behavior, leading to correct schedules for such cases.
(2) Refining control relations to be more granular.
We plan to generalize the resource model to capture a wider range of resource types, including GPU memory, network bandwidth, and heterogeneous memory hierarchies, as well as to support dynamic resource profiling during execution. In this paper, we only consider CPU-compute functions and consider two such function calls to be mutually exclusive if they are assigned the same rank.
Yet, this strict criterion can lead to increased latency for GPU-compute functions, as two of such calls could still be coordinated for parallel execution if they do not occupy the full VRAM.
We believe that making mutual-exclusive relations much finer-grained in the future could bring further speedups.

\smalltitle{Threats to Validity}
The first threat is the potential bias of the intentionally crafted (KITTI$^{\ast}$) benchmarks, which may not fully represent practical real-world scenarios and could be influenced by human preferences.
We tried our best to sample diverse and representative data and create queries simulating real-world scenarios.
The second threat is that the benchmarks we employed in our evaluation might not be as complex as real-world scenarios.
To mitigate this threat, we initially used the same benchmarks as \llmcompiler, the state-of-the-art baseline in parallel function orchestration.
Additionally, we designed additional benchmarks (KITTI$^{\ast}$ and AGNews$^{\ast}$), ensuring that each query included a minimum of eight intensive compute calls, each lasting for at least $\ge$65 seconds when executing on M1 Pro, along with several input calls.
Furthermore, we provided case studies involving real-world scenarios to assess the practicality of \ourapproach.
We believe that such experimental settings are sufficient to evaluate the performance of \ourapproach.

\smalltitle{Implications for Software Engineering}
{Beyond latency improvements, \ourapproach contributes to the software engineering of LLM-driven systems by providing an explicit architectural pattern. By representing workflows as a Function Relation Graph (FRG), it enforces separation of concerns, enabling predictable and analyzable behavior instead of opaque LLM reasoning. This explicit representation enhances maintainability, allowing developers to add tools modularly and assess downstream impacts. Moreover, \ourapproach incorporates principled fault-tolerance, replacing ad-hoc error handling with systematic recovery, and leverages resource-aware scheduling to ensure stable operation. In this way, \ourapproach serves as a blueprint for building reliable, evolvable, and maintainable LLM-based software, aligning established SE principles with the emerging domain of AI agents.}

\section{Related Work}\label{sec:related_work}

\smalltitle{Function Orchestration}
Function calling has been a fundamental capability of LLMs since the advent of ReAct~\cite{yao2022react}, to the best of our knowledge.
ReAct orchestrates function calls by sequentially generating a thought, selecting a function, and observing the function's outcome for the next iteration unless the user query is resolved.
Subsequently, frameworks like LangChain~\cite{Chase_LangChain_2022} followed ReAct to orchestrate functions sequentially and interact with LLMs.
On June 13, 2023, OpenAI enabled native function calling for its GPT-series models~\cite{openaifunccall}, similar to ReAct.
After that, datasets~\cite{glaiveai,patil2023gorilla}, tools~\cite{qin2023toolllm}, and LLM variants~\cite{functionary,patil2023gorilla} were also created to augment existing open-source LLMs with native function calling capabilities.
Berkeley created a leader board to benchmark LLM's function calling capabilities~\cite{berkeleyfcleaderboard}.
Today, function calling has been a defacto capability for LLMs like LLaMA3~\cite{LLaMa3}, ChatGLM3~\cite{chatglm3}, and GLM4~\cite{glm4}.

As of November 6, 2023, OpenAI upgraded the function calling capability to Parallel Function Calling~\cite{openaipfcandex1,openaipfcrel,openaipfcpr}, enabling LLMs to produce multiple function call requests at a time. 
This enhancement aims to improve efficiency when orchestrating functions.
However, such capabilities currently only exist in GPT-series models.
Meanwhile, \llmcompiler proposed a framework to orchestrate function calls concurrently for any LLMs~\cite{kim2024an}.
Like \ourapproach, \llmcompiler translates a user query into a sequence of function calls, analyzes their inter-dependencies, and executes all concurrent function calls directly.
\llmtoolcompiler optimizes function calling by analyzing and fusing multiple calls into a single one~\cite{singh2024llm}.

In contrast, \ourapproach concentrates on discovering and leveraging the data and control relations among function calls to improve the efficiency.

\smalltitle{Prompt Engineering}
Another series of work is to devise efficient approaches to interacting with LLMs and maximizing their inherent, inference capabilities~\cite{liu2023prompting}.
These works can be categorized into three groups:
(1) Prompting Approaches for example zero- and few-shot prompting, chain~\cite{wei2022chain} or tree~\cite{tot} or graph~\cite{got} of thoughts, and retrieval augmented generation~\cite{lewis2021retrievalaugmented}.
(2) Prompting Languages for simplifying the interactions with LLMs, such as LMQL~\cite{beurer2023prompting}, SGLang~\cite{zheng2023efficiently}.
3) LLM Wrappers which integrates LLMs into conventional programming languages or tools, such as Semantic Kernel~\cite{Semantic-kernel}, LangChain~\cite{Chase_LangChain_2022}, SuperAGI \cite{SuperAGI}, and NeMo-Guardrails~\cite{NeMo-Guardrails}.
These works are orthogonal to ours and can be integrated into \ourapproach--for example query translation--for better controlling LLMs.

\smalltitle{Low-Level Optimizations}
Some works focus on accelerating LLM's training, fine-tuning, and inference~\cite{lowlevelsurvey}.
Unlike us which put efforts in function calling, these works typically target token-~\cite{llmstream}, data-~\cite{llmlingua}, model-~\cite{gu2024mamba}, or system-level~\cite{kwon2023efficient} optimizations or acceleration.
They can be combined with us.

\smalltitle{Multithreaded Computation}
Efficiently scheduling concurrent tasks to maximize resource utilization and throughput is crucial.
To our knowledge, Early List Scheduling~\cite{list_schedule} established the theoretical foundation for parallel scheduling.
It makes static scheduling decisions based on estimated task times and dependencies, but lacks dynamic adjustment, limiting its effectiveness in load balancing and dependency management.
Work Stealing~\cite{workstealing,288650} addressed the load imbalance issue of List Scheduling, by allowing idle processors to steal tasks from busy ones. But it brought the task distribution and migration, resulting in complex task dependencies, and scalability issues.
Classic strategies like Shortest Job First (SJF)~\cite{sjf}, First In-First Out(FIFO)~\cite{FIFO-TSE} and Earliest Deadline First (EDF)~\cite{10.5555/552538} prioritize tasks based on the shortest execution time and earliest deadline, respectively.
These methods require prior knowledge or estimation of task execution times.
Additionally, some approaches~\cite{5214359, 10.5555/2002181.2002183} optimize parallel scheduling specifically for GPUs, focusing on dense linear algebra computations, while others~\cite{8855446} aim to be suitable for heterogeneous environments (GPU/CPU).
\ourapproach's approach follows a work-sharing~\cite{workstealing} style, yet it differentiates between two types of relations.
\section{Conclusion}\label{sec:conclusion}
We present \ourapproach, an advanced framework for parallel function calling in large language models. 
The main idea behind \ourapproach is to identify an available processor to execute a function call while ensuring that no single processor is overwhelmed.
To achieve this, \ourapproach models the data relations among function calls and coordinates their execution based on control relations and the work status of processors.
When comparing against state-of-the-art techniques, \ourapproach showed similar efficiency improvements for inout function calls and significantly outperformed them for compute calls.
\ourapproach also demonstrated potential in real-world scenarios.

\section*{Acknowledgment}
We thank the anonymous reviewers for their constructive comments to improve our article.
This research is supported by the Ministry of Education, Singapore under its Academic Research Fund Tier 2 (Award ID: T2EP20222-0037) and the Key R\&D Program of Zhejiang under Grant No. 2025C01083. This work was supported by Ant Group.

\bibliographystyle{IEEEtran}
\bibliography{main}

\end{document}